\documentclass[sigconf, nonacm]{acmart}
\pdfoutput=1
\usepackage{array}
\usepackage{csquotes}
\usepackage{soul}
\usepackage{todonotes}
\usepackage{listings}
\usepackage{longtable}
\usepackage{tabularx}
\usepackage{geometry}
\usepackage[utf8]{inputenc}
\usepackage{enumitem}
\usepackage{multirow}
\usepackage{tabularx}
\usepackage{longtable}
\usepackage{array}
\usepackage{graphicx}
\usepackage{makecell} 
\usepackage{supertabular}
\usepackage{afterpage}
\RequirePackage{color}\definecolor{RED}{rgb}{1,0,0}\definecolor{BLUE}{rgb}{0,0,1} 

\copyrightyear{2025}
\acmYear{2025}
\setcopyright{cc}
\setcctype{by}
\acmConference[CHI ’25]{CHI Conference on Human Factors in Computing Systems}{April 26-May 1, 2025}{Yokohama, Japan}
\acmBooktitle{CHI Conference on Human Factors in Computing Systems (CHI ’25), April 26-May 1, 2025, Yokohama, Japan}\acmDOI{10.1145/3706598.3713701}
\acmISBN{979-8-4007-1394-1/25/04}

\makeatletter
\patchcmd{\csq@bquote@i}{{#6}}{{\emph{#6}}}{}{}
\makeatother

\begin{document}

\title[Rescriber: Smaller-LLM-Powered User-Led Data Minimization for LLM-Based Chatbots]{Rescriber: Smaller-LLM-Powered User-Led Data Minimization for LLM-Based Chatbots}

\author{Jijie Zhou}
\affiliation{%
  \institution{Northeastern University}
  \city{Boston}
  \country{USA}
}

\author{Eryue Xu}
\affiliation{%
  \institution{Northeastern University}
  \city{Boston}
  \country{USA}
}

\author{Yaoyao Wu}
\affiliation{%
  \institution{Northeastern University}
  \city{Boston}
  \country{USA}
}

\author{Tianshi Li}
\affiliation{%
  \institution{Northeastern University}
  \city{Boston}
  \country{USA}
}

\renewcommand{\shortauthors}{Zhou et al.}
\newcommand{\tlcomment}[1]{\noindent{\textcolor{blue}{\textbf{\#\#\# TL:}\textsf{#1} \#\#\#}}}
\newcommand{\jzcomment}[1]{\noindent{\textcolor{orange}{\textbf{\#\#\# JZ:}\textsf{#1} \#\#\#}}}
\newcommand{\samplesize}{12}
\newcommand{\qualifiedsamplesize}{105}
\newcommand{\totalsamplesize}{275}
\newcommand{\tookminimizationactionsize}{11}
\newcommand{\systemname}{Rescriber}
\definecolor{lightblue}{HTML}{c4e6ff}
\definecolor{lightred}{HTML}{ffe1e1}

\newcommand{\modelone}{\colorbox{lightred}{\texttt{\systemname{}-GPT-4o}}}
\newcommand{\modeltwo}{\colorbox{lightblue}{\texttt{\systemname{}-Llama3-8B}}}
\newcommand{\modeloneplain}{\texttt{\systemname{}-GPT-4o}}
\newcommand{\modeltwoplain}{\texttt{\systemname{}-Llama3-8B}}
\newcommand{\modeloneintable}{\systemname{}-GPT-4o}
\newcommand{\modeltwointable}{\systemname{}-Llama3-8B}

\renewcommand{\sectionautorefname}{Section}
\renewcommand{\subsectionautorefname}{Section}
\renewcommand{\subsubsectionautorefname}{Section}

\definecolor{myblue}{RGB}{72, 120, 208}
\definecolor{myorange}{RGB}{238, 133, 74}
\definecolor{mygreen}{RGB}{106, 204, 100}
\definecolor{myred}{RGB}{214, 95, 95}
\definecolor{mypurple}{RGB}{149, 108, 180}
\definecolor{mycyan}{RGB}{130, 198, 226}
\definecolor{mypink}{RGB}{220, 126, 192}
\definecolor{mybrown}{RGB}{140, 97, 60}

\definecolor{gray}{gray}{0.5}

\begin{abstract}
The proliferation of LLM-based conversational agents has resulted in excessive disclosure of identifiable or sensitive information.
However, existing technologies fail to offer perceptible control or account for users' personal preferences about privacy-utility tradeoffs due to the lack of user involvement.
To bridge this gap, we designed, built, and evaluated \systemname{}, a browser extension that supports user-led data minimization in LLM-based conversational agents by helping users detect and sanitize personal information in their prompts.
Our studies (N=\samplesize{}) showed that \systemname{} helped users reduce unnecessary disclosure and addressed their privacy concerns.
Users' subjective perceptions of the system powered by Llama3-8B were on par with that by GPT-4o.
The comprehensiveness and consistency of the detection and sanitization emerge as essential factors that affect users' trust and perceived protection.
Our findings confirm the viability of smaller-LLM-powered, user-facing, on-device privacy controls, presenting a promising approach to address the privacy and trust challenges of AI.

\end{abstract}



\maketitle

\section{Introduction}
\begin{figure*}[t]
    \centering
    \includegraphics[viewport=0 50 1950 1040,width=1\textwidth, clip]{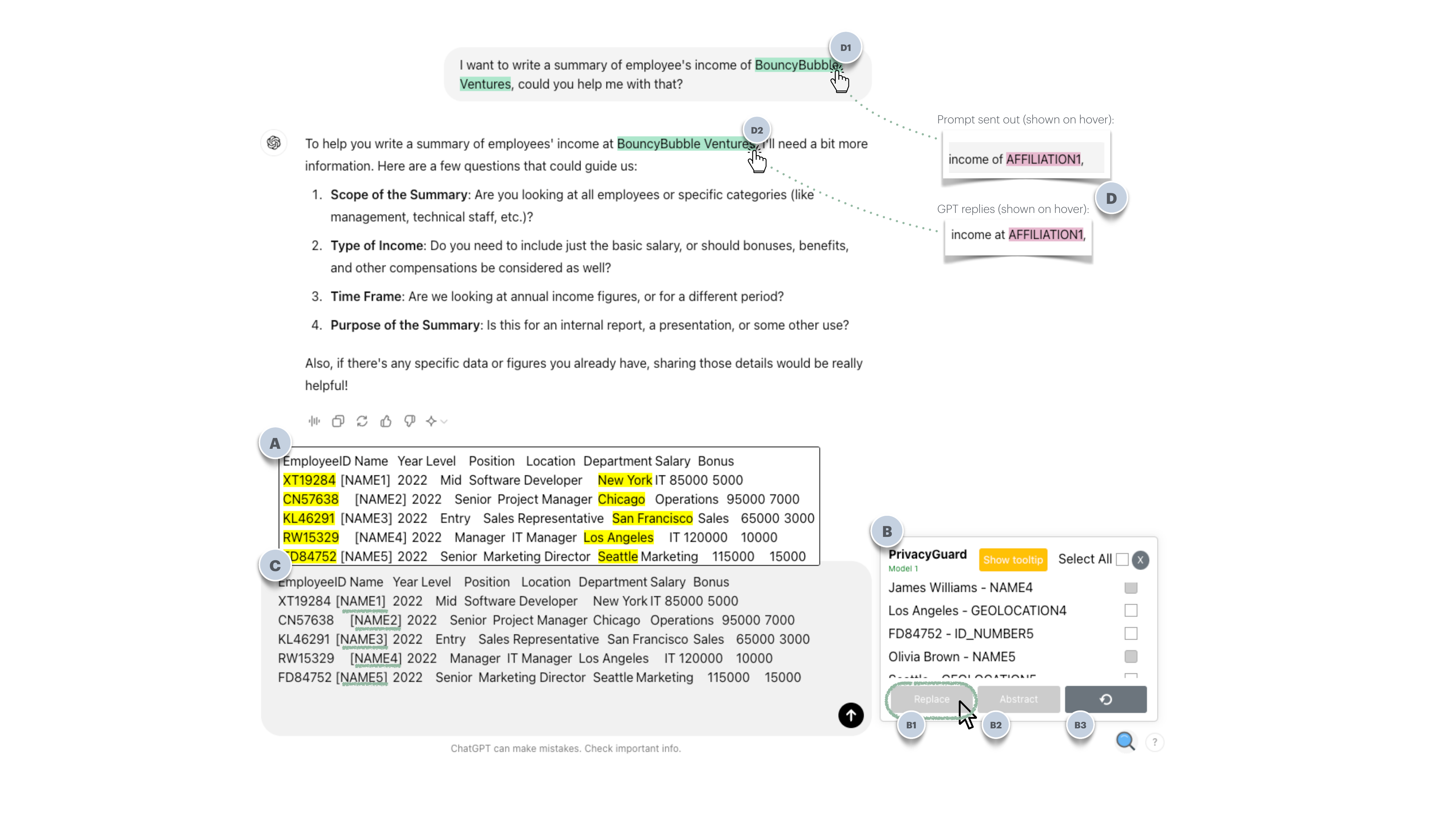}
    \caption{A snapshot of \systemname{} user experience when using ChatGPT for data analysis. 
    \systemname{} displays a tooltip that highlights the detected personal information in the user's message (A), and offers a control panel (B) where users can either replace the information with a placeholder (B1), abstract the information to a more general version (B2), or revert the actions (B3). All the changes would be immediately effective on the message that would be sent out in the input box (C). In this example, the user replaced the detected names (e.g., James Williams) to corresponding placeholders (e.g., [NAME4]).
    For the sent messages (D1) and ChatGPT's responses (D2), \systemname{} replaces the placeholders with the original PIIs to increase readability and facilitate copying. When the user mouses over the highlighted PIIs, the placeholders used in the actual messages sent and received are revealed (D).}
    \label{fig:teaser}
    \Description{Rescriber User Interface for Handling Personal Information in ChatGPT Messages. This figure shows the interface of Rescriber, which is used to handle personal data in ChatGPT conversations. The user's input (A) contains several pieces of personal information (such as names, locations, and salaries), highlighted by the system. A control panel (B) allows users to manage this information by either replacing the personal data with placeholders (B1) or abstracting it into a more general format (B2). Users can also revert the modifications (B3). The placeholders (e.g., "NAME4," "ID_NUMBERS") replace the specific personal data in both the user’s message (D1) and ChatGPT’s responses (D2). When users hover over the placeholders, the original PII is revealed.}
\end{figure*}

The emergence of large language models like OpenAI's GPT~\cite{openaichatgpt}, Google's Gemini~\cite{googlegemini}, and Meta's Llama~\cite{touvron2023llama} has enabled LLM-based conversational agents that understand the context and can generate human-like responses.
As products with high capacity for handling natural language tasks and user-friendly interfaces, such as ChatGPT, come into play, people increasingly rely on them for a variety of tasks in personal and professional contexts~\cite{Zhang2023ItsAF, cheng2023now, bom2023exploring}.

Consequently, a large amount of personal or sensitive disclosure has been observed in users' conversations with ChatGPT-like LLM-based conversational agents~\cite{Zhang2023ItsAF, mireshghallah2024trustbotdiscoveringpersonal}, which leads to severe privacy threats in the event of database breaches and data extraction attacks targeting the memorization vulnerabilities of LLMs~\cite{carlini2021extracting, carlini2022quantifying, zhang2021counterfactual, nasr2023scalable}.
This includes personally identifiable information (PII) like names and email addresses, as well as non-PII sensitive topics such as sexual preferences and drug use~\cite{mireshghallah2024trustbotdiscoveringpersonal}.
Notably, parts of the disclosure are unnecessary to the main task, like
hyper-detailed information: ``Help me generate a one-day itinerary in NYC, I live at \textit{153 W 57th St, New York, NY 10019}'', or irrelevant information: ``Please help me proofread the following email to my colleague \textit{peter} (\textit{peter.parker@spider.com})''.

The excessive and sensitive disclosure issues pose a challenge to the classic data privacy principle, \textit{data minimization}~\cite{schaar2010privacy, gdpr2016, fipps}.
Although researchers and practitioners have employed data sanitization~\cite{Peris_Dupuy_Majmudar_Parikh_Smaili_Zemel_Gupta_202, dou2023reducing, olstad2023generation} and private execution in trusted hardware~\cite{privatecloudcompute, tramer2018slalom} to achieve the goal of removing PII at the training stage and avoiding unnecessary data retention at the inference stage, there are several inherent limitations of the techniques due to the lack of involvement of the end-users.
First, the privacy and utility needs vary by context and person, meaning post-hoc data sanitization may not remove all the information the user found sensitive or unnecessary.
Second, once data leaves the device, users lose direct control over their information.
Existing techniques operate in a concealed way, making it difficult for users to form accurate mental models, reduce privacy concerns, and trust LLM-based agents with sensitive information.
These issues reveal a research gap in designing user-friendly tools that enhance users' ability to navigate privacy risks and minimize unnecessary information disclosure, specifically enabling \textit{user-led data minimization}.

Prior research has emphasized the trade-offs among privacy, utility, and convenience as a major challenge to achieving privacy-preserving disclosure behaviors to LLM-based conversational agents~\cite{Zhang2023ItsAF}.
This is especially true when it comes to the excessive disclosure issue.
Excessive disclosure often arises when users paste texts with semantically unrelated information they didn't notice or find too cumbersome to remove.
\citet{Zhang2023ItsAF} found that users sometimes manually sanitize inputs, but the manual actions fall short due to forgetfulness and tediousness.
They also found that users were pessimistic about preserving privacy while reaping the benefits of AI, and users were neither capable of identifying all the opportunities associated with minimizing excessive sensitive disclosures nor willing to invest in efforts and time to do that~\cite{Zhang2023ItsAF}.
Hence, we argue that a tool that supports user-led data minimization should act in a way that \textit{maintains response quality} and \textit{incurs minimal overhead}.

In this work, we make an initial foray into designing user-led data minimization support for LLM-based conversational agents to tackle the aforementioned challenges.
Specifically, we aim to design a browser extension that leverages AI to streamline the process of identifying and sanitizing unnecessary disclosures of personal information, as LLMs significantly outperform traditional NLP methods for detecting PII \cite{bubeck2023sparks}.
There are several key challenges in achieving this goal.
First, accurate detection and sanitization recommendations are essential for effective support for user-led data minimization~\cite{zhang2024designing}.
According to the neural scaling laws~\cite{wei2022emergent, schaeffer2024emergent}, using larger models increases the accuracy, while it also imposes higher computational costs and raises the bar of the user's device for deploying the system locally.
Furthermore, it's difficult to determine the sweet spots of leveraging large language models to achieve the ultimate goal of empowering users to staying in control over their data while maintaining the response utility.
Lastly, it is unclear how different approaches to data minimization affect users' privacy perceptions and behaviors.  
Formally, we aim to investigate the following research questions on core efficacy, adoption intention, and impact on users:

\begin{itemize}
    \item[RQ1] How can we design a tool to support effective data minimization and reduce users' privacy concerns?  
    \item[RQ2] How can we design a tool for user-led data minimization that users trust and intend to use? 
    \item[RQ3] How does the support for user-led data minimization affect users' privacy awareness and behaviors? 
\end{itemize}

We present \systemname{}, a browser extension for Chrome, which detects and highlights potential personal information disclosures, and then provides users with options to redact or abstract these details before sending a message.
Redaction replaces personal terms with entity placeholders (e.g., ``Alex'' becomes ``[NAME1]''), while abstraction replaces them with generalized and contextually appropriate information (e.g., ``my colleague'').
When the address is highlighted as personal information in ``Help me generate a one-day itinerary in NYC, I live at 153 W 57th St, New York, NY 10019'', we could abstract the information and the message would turn into ``Help me generate a one-day itinerary in NYC, I live near Central Park'', this way, the contextual information still exists, but with a level of details that would not expose too much of the user's private info.
Likewise, when the email address ``peter.parker@spider.com'' is highlighted and suggested by the tool to be replaced, users could directly redact it with a placeholder like ``[EMAIL1]'' since it would not change the contextual meaning and exposing the specific contact info won't necessarily help generate a better response.
This approach endows users with the flexibility of modifying the message to preserve privacy while retaining the essential information to maintain utility simultaneously.
\autoref{fig:teaser} depicts the framework of the proposed extension, illustrating example actions and outcomes using \systemname{}.

To address our research questions, we conducted usability studies with \samplesize{} ChatGPT users on two prototype versions: \modeltwoplain{} and \modeloneplain{}.
The former is powered by a smaller LLM that runs on a consumer device (e.g., a Mac Mini with an Apple M-series chip), while the latter serves as a reference point for state-of-the-art LLMs that require massive computing resources and cannot be hosted locally.
Each participant tested both versions to sanitize a hypothetical prompt provided by us and a real-world ChatGPT prompt, providing feedback on the perceived privacy protection and satisfaction with the sanitized response.
We found that both versions helped users reduce excessive disclosure of personal information in all tested prompts.
The \modeloneplain{} helped reduce more unnecessary disclosure than \modeltwoplain{} due to the more comprehensive detection of the larger model.
However, users' subjective preferences for the two models were similar in terms of data minimization support and adoption intention.

Users' main concerns with ChatGPT stem from worries about being identifiable and the potential risks due to the lack of transparency and sense of control over how their data is handled once shared, both of which can be effectively mitigated by our tool.
Participants appreciated the tool’s features for sanitizing messages without disrupting their workflow, giving them control over sanitization actions, and offering convenient support for privacy preservation without compromising response quality.
They considered the comprehensiveness and consistency of detection and sanitization as key trust factors, emphasizing recall over precision.
In addition to the benefits of data minimization, participants appreciated the educational value of our tools, as the detection feature increases their awareness of sensitive information disclosure and elicits reflections on the potential risks.
Our results verified the feasibility of using off-the-shelf, smaller LLMs to provide on-device, user-facing data minimization support and offer concrete design recommendations for using smaller LLMs as a middle layer for privacy-preserving access to larger LLMs.
In summary, our key contributions include:
\begin{itemize}
\item[1] Designing and implementing \systemname{}, a browser extension that leverages on-device LLM to facilitate user-led data minimization in LLM-based conversational agents\footnote{Source code: \url{https://github.com/PEACH-Research-Lab/Rescriber_frontend_ondevice}}.
\item[2] Conducting usability studies to evaluate \systemname{} in terms of the efficacy of data minimization support, users' perceived privacy protection, adoption intention, and impact on users.
\item[3] Proposing the concept of smaller-LLM-powered, user-facing, on-device privacy controls and synthesizing concrete design recommendations based on a thorough mixed-methods analysis of our user studies.
\end{itemize}

\section{Related Work}

\subsection{Privacy issues in LLMs and LLM-based Conversational Agents}

LLM-based conversational agents (CAs) present two main types of privacy threats~\cite{Zhang2023ItsAF}. The first involves traditional security and privacy risks, such as data breaches and the sale of personal information. The process of user interaction with LLM-based CAs are also vulnerable to cyberattacks, data leaks, or ransomware threats~\cite{kshetri2023cybercrime}.
The second type of threat is specific to LLMs' memorization risks~\cite{carlini2021extracting, carlini2022quantifying, zhang2021counterfactual, nasr2023scalable}.
Research has shown that LLMs can inadvertently retain and reproduce precise details from their training datasets.
The memorization issue poses significant privacy and security concerns, especially when the training data includes sensitive or personal information.
The cost of exploiting this vulnerability is relatively low.
For example, research has found that by prompting the model to continuously output ``poem,'' it can be manipulated into revealing its training data verbatim~\cite{nasr2023scalable}.
When individuals use LLM-based CAs, they often share an excessive amount of personal information~\cite{mireshghallah2024trust, Zhang2023ItsAF}.
Given that the user prompts are often used for model training or fine-tuning, and users are often unaware of the opt-out options~\cite{Zhang2023ItsAF}, the excessive disclosure behaviors are subject to both of the aforementioned threats.
Note that there is also research discussing the privacy issues of using LLMs to infer sensitive attributes~\cite{staab2023beyond, mendes2024granular}, as well as unintentional privacy leakage issues of LLM-facilitated interpersonal communication~\cite{mireshghallah2023can, shao2024privacylens}, while they are different from the threat model that we focus on in this work, which is the privacy concerns regarding sharing information with the LLM service provider.

Existing privacy mitigation methods for the two types of threats focus on various stages of the model's lifecycle, including pre-training, post-training, and inference. Prior work has explored privacy-preserving techniques, especially in addressing the memorization risk, from a model-centric perspective. During model training phase, techniques such as data sanitization~\cite{lison2021anonymisation,kandpal2022deduplicating} and differentially private training methods~\cite{li2021large,yu2021differentially} are commonly employed.
Post-training, approaches like ``knowledge unlearning''~\cite{jang2022knowledge} remove knowledge tied to specific token sequences.
At inference, privacy risks can be mitigated using PII detection and differentially private decoding~\cite{majmudar2022differentially}.
However, these methods still have limitations and are debated. For example, although fine-tuning models to selectively ``unlearn'' content has been proposed, sensitive details closely related to the series can still be recalled in various ways~\cite{shostack2024boy}. 

Current model-centric mitigation strategies do not fully address the privacy threats and concerns associated with interacting with LLM-based CAs.
In this work, we study a user-driven mitigation approach which aims at enhancing users' control over their input and build trust in privacy protection.
Users often share excessive information when using these agents~\cite{mireshghallah2024trust, Zhang2023ItsAF}, compromising privacy for convenience~\cite{li2024humancenteredprivacyresearchage}.
The design of our system, \systemname{}, was informed by these issues to emphasize helping users manage the trade-offs among privacy, utility, and convenience in a way that aligns with their interests.
We envision our method as orthogonal and complementary to the model-centric mitigation approaches.

\subsection{Data minimization}

Data minimization is a fundamental privacy protection principle, which is reflected in several foundational privacy frameworks and privacy laws. According to ``Privacy by Design,'' data processing systems should be designed and selected to collect, process, and use no personal data , or as little personal data as possible~\cite{schaar2010privacy}. Similarly, the Fair Information Practice Principles (FIPPS) stipulate that agencies should collect or disclose only directly relevant and necessary PII for accomplishing a legally authorized purpose, and should only retain PII for as long as necessary to fulfill that purpose~\cite{fipps}.
This emphasis on data minimization is also evident in legal frameworks such as the European Union’s General Data Protection Regulation (GDPR). Article 5(1)(c) of the GDPR mandates that personal data must be limited to what is necessary for processing purposes, thus reinforcing the principle of data minimization~\cite{gdpr2016}.

There has been a lot work on supporting and understanding how to implement data minimization in the privacy design and engineering process.
One line of work focuses on auditing the compliance of the data minimization principle.
For instance, \citet{li2017privacystreams} presents PrivacyStreams, a functional programming model that helps mobile developers collect data with specified granularity, and simplifies auditing.
Other works are more driven by legal compliance requirements.
For example, \citet{rastegarpanah2021auditing} proposes a method for auditing black-box prediction models for compliance with the GDPR’s data minimization principle.
\citet{biega2020operationalizing} explores how to operationalize data minimization for personalization, which inherently conflicts with privacy goals, by asking: How much information and what information does an individual need to provide to receive quality personalized results?
On the other hand, research shows developers had trouble implementing data minimization due to uncertainty of the amount information that potentially needs to be collected at the design phase~\cite{senarath2018understanding} and a tendency to collect data for future analysis~\cite{li2018coconut}.

Given the challenges of implementing data minimization solely during the software design phase, it is increasingly important to also involve users in enacting this principle.
For example, \citet{sharma2024m} studied how data minimization could be designed in search engines from a user-centered perspective.
Building on this idea, our work focuses not on redesigning the entire system, but on proposing a practical approach that enhances user-facing data minimization support. This approach gives users more flexible control over data sharing and raises their awareness of the risks and opportunities in minimizing unnecessary disclosures.

\subsection{User Agency in Privacy Protection}

``Privacy as control'' is an essential conception of privacy that has deeply influenced privacy laws in many regions, including the U.S. and the EU~\cite{waldman2018privacy}.
In usable privacy research, designing user-friendly privacy control features to increase users' agency in privacy protection has been a major focus.
\citet{sharma2024m} explores user-controlled data minimization in search engines, highlighting the importance of end users in managing their own data privacy. It shows that enabling user agency through customizable and transparent data minimization features can significantly enhance privacy protection in search engines. 
Many studies demonstrate an evolution from simple manual controls to sophisticated, user-centric approaches in privacy designs, incorporating tangible solutions, perceptible assurance, AI-facilitated methods and accessibility considerations.
\citet{do2021smart} introduced a Smart Webcam Cover for tangible, assisted privacy control.
Smart speakers implemented wake-up words as a user-initiated mechanism.
\citet{do2023powering} developed a wireless microphone powered by intentional user interactions, providing perceptible assurance.
\citet{zhang2024designing} explored AI-facilitated data sanitization for photo sharing among visually impaired users.
Notably, many of these approaches primarily offer coarse-grained, binary control over data collection, limiting users' ability to control their privacy preferences.

Our work investigates the challenges of increasing user agency in privacy specifically in the context of LLM-based CAs, such as ChatGPT.
These applications posed unique challenges to privacy agency due to the intense tensions among privacy, utility, and convenience~\cite{Zhang2023ItsAF}.
To address this, we leverage large language models to offer a user-centered approach to data protection, providing users with granular, real-time privacy control.
\citet{chong2024casper} developed a system that also aimed at prompt sanitization using web-based LLMs.
Their work focused on detection and sanitization techniques, while our work further contributes to the design of a user-facing tool and user studies that provide in-depth insights into the design considerations, as well as measuring the end-to-end effectiveness of users-led data minimization.

\section{The \systemname{} System}
\begin{figure*}[t]
    \centering
    \includegraphics[viewport=0 300 1950 1000,width=1\textwidth, clip]{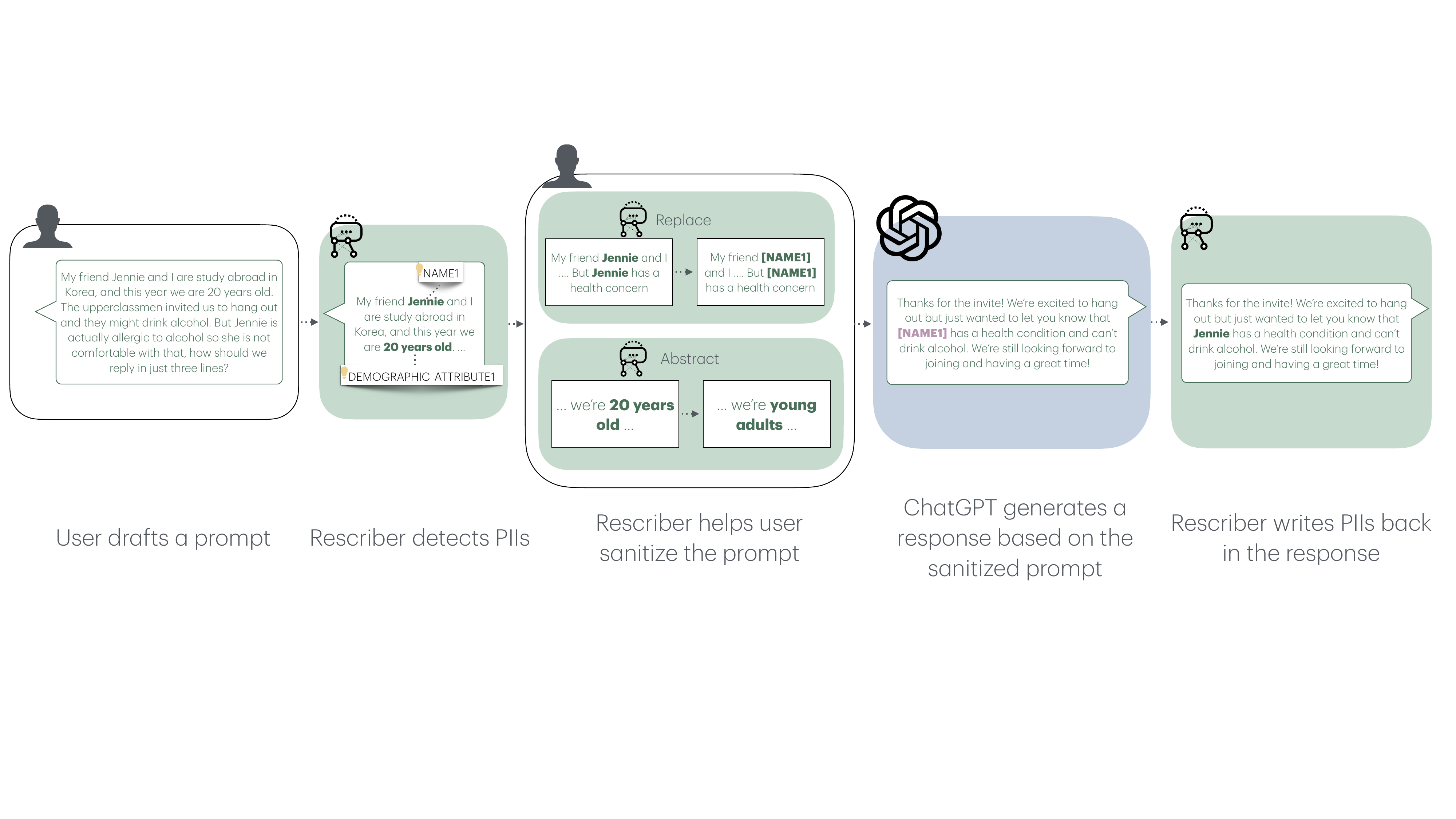}
    \caption{Main stages and features of \systemname{}: 1) User types the message in ChatGPT's prompt entry box; 2) \systemname{} automatically detects and highlights the sensitive information; 3) user can redact their message based on \systemname{}'s suggestion. Once 4) GPT generates a response based on the sanitized message, and if GPT uses the part that user chooses to redact, 5) \systemname{} will help replace the part back to the original information for better utility.}

    \label{fig:architecture}
    \Description{Rescriber Process Flow for Prompt Sanitization. This figure shows the main stages of the Rescriber system, which helps users redact personal information before sending a prompt to ChatGPT. The process begins with the user drafting a prompt, which \systemname{} then analyzes to detect personal information. It suggests replacements or abstractions for any sensitive data identified. Once the prompt is sanitized, ChatGPT generates a response. Finally, \systemname{} reintroduces the redacted information back into the response to maintain context and utility.}
\end{figure*}

\subsection{Overview}
We developed \systemname{}, a Chrome extension prototype, to help users minimize data when interacting with LLM-based conversational agents like ChatGPT.
The tool makes privacy-related data minimization visible without interfering with users' regular GPT usage.
Figure~\ref{fig:architecture} illustrates the system's workflow and its role during user interactions with GPT.

\subsubsection{Design Goals}
The design and main functionality of \systemname{} are guided by three design goals, as detailed below.

\paragraph{D1: Perceptible privacy protection} \label{d1} Existing solutions to privacy risks of LLMs~\cite{Peris_Dupuy_Majmudar_Parikh_Smaili_Zemel_Gupta_202, dou2023reducing, olstad2023generation, privatecloudcompute} face challenges in building trust with users, as their effects are often invisible and users lack assurance about whether they prevent data leakage.
To address this, we try to resolve it at its source by using a ``what-you-see-is-what-you-get'' model.
This allows users to alleviate their privacy concerns before sending data to GPT, giving them perceptible safeguards over the use of ChatGPT-like LLM-based conversational agents.

\paragraph{D2: Protecting privacy without compromising utility and convenience}
\label{d2} We want users to adopt the tool and develop a habit of using privacy-preserving techniques, while ensuring that the utility of GPT responses remains intact.
Additionally, the tool should sanitize messages without compromising response effectiveness or adding extra burden, as convenience is a key factor in ChatGPT use and often leads to excessive data disclosure~\cite{Zhang2023ItsAF}.

\paragraph{D3: Flexible control over information sanitization}
\label{d3}
Privacy concerns and preferences are nuanced, context-sensitive~\cite{nissenbaum2004privacy}, and vary from person to person~\cite{malhotra2004internet}.
Therefore, our tool must offer rich and flexible support in reducing the specificity and amount of information shared.
The flexibility enhances users' agency and builds users' trust in the tool.
Additionally, providing granular control also helps users better minimize unnecessary disclosure by striking a balance between privacy and utility.

\subsubsection{Threat model}
Disclosure of sensitive personal information to LLM-based conversational agents can lead to various threats to privacy including conventional data breach risks and AI-related risks in the model training, fine-tuning, and inference stages.
Commercial chatbots may use the user chat data for model training or fine-tuning.
The model may be trained to memorize user input and can be prompted to regurgitate sensitive information to unauthorized users~\cite{carlini2022quantifying, carlini2021extracting}.
At the inference stage, commercial chatbots may extract personal information from the chat data to develop a user profile for personalized response generation (e.g., ChatGPT's memory feature).
This could inadvertently include personal data in responses, risking sensitive information leaks~\cite{shao2024privacylens, zhang2024can}.
Rescriber aims to assist users in being more mindful of their disclosure to LLM-based conversational agents and adhering to the data minimization principle, hence helping mitigate the above privacy threats and maintain an effective balance between privacy and utility.

\subsection{Example Usage Scenario}
We present a hypothetical example to illustrate a typical use case of \systemname{} (also illustrated in \autoref{fig:architecture}).
Nova is a college student studying abroad in South Korea. She and her friend got invited by some upperclassmen to a party where drinking might be involved, but her friend Jennie cannot consume alcohol. With limited knowledge of how locals handle these situations, Nova turns to GPT and types a description, explaining that she and Jennie, both 20, are of legal drinking age in Korea (19) while Jennie is allergic to alcohol, making it tricky for them as newcomers. Before sending, \systemname{} automatically flags personal and sensitive information, such as the name, ages, and Jennie's health condition. 
Initially unaware of how much identifiable information was included, Nova reviews the flagged suggestions and agrees that certain details should be anonymized. She replaces ``Jennie'' with a placeholder \systemname{} provided (``NAME1'') and let \systemname{} abstract ``20 years old'' to a more general term, ``young adults''. She keeps the reference to Korea, for cultural context, and retains the mention of Jennie’s alcohol allergy, as it is essential. 
GPT responds using the placeholders Nova had used in the sanitized message. To her pleasant surprise, \systemname{} automatically replaces the placeholder with the original terms (e.g. ``Jennie'') and highlights them for readability. Nova is happy with the simple process that protected her privacy and provided a ready-to-use response.

\subsection{Detection} \label{detection}
When the user stops typing, the model detects PII or personal experiences in the message and highlights them to prompt further sanitization. This real-time feature lets users review their message before sending (\hyperref[d1]{\textbf{D1}}).

\subsubsection{Process}
Detection is achieved by prompting an LLM using a pre-defined taxonomy of personal information that could potentially be identifiable or sensitive, as detailed in \autoref{Taxonomy and Prompt Design}.
Results are formatted as a JSON object, listing all the detected entities and their data type categories (e.g., NAME), which are used to create placeholders for future replacements.
If multiple entities of the same category exist in a conversation, the system appends a number to each placeholder (e.g., NAME1). 
The mapping of entities to placeholders is stored locally, and users maintain full control over this data, with the option to delete it at any time.
A magnifier icon signals successful detection at the bottom right of the webpage. Clicking it lets users view highlighted entities in a tooltip or review them with placeholders in a control panel.

\subsubsection{Taxonomy and Prompt Design} \label{Taxonomy and Prompt Design}

The prompt and taxonomy used for detection were based on \citet{bubeck2023sparks} and the self-disclosure categories from \citet{dou2023reducing}, with two main enhancements.
First, we introduced the instruction to exclude public information matching a certain PII entity from detection. For example, celebrity names (e.g., Taylor Swift) are generally not considered private and should not be flagged.
However, they should be detected when used in private contexts (e.g., Sam Altman sends an email to his colleague).
Therefore, our instruction defines the criteria as \textit{``information should be related to a real person not in a public context, but okay if not uniquely identifiable''}.
Another enhancement is that we included more data types about sensitive personal experiences in addition to the conventional PII types in our taxonomy (\autoref{tab:taxonomy}).
We include health information, financial information, and educational records, as they are crucial data types regulated by sectoral consumer privacy laws in the U.S.
Below is the system prompt.

\textit{You are an expert in cybersecurity and data privacy. You are now tasked to detect PII from the given text, using the following taxonomy only: \{taxonomy omitted\}}

\textit{For the given message that a user sends to a chatbot, identify all the personally identifiable information using the above taxonomy only, and the entity\_type should be selected from the all-caps categories.}

\textit{Note that the information should be related to a real person not in a public context, but okay if not uniquely identifiable.
Result should be in its minimum possible unit.
Return me ONLY a json in the following format: \{"results": [\{"entity\_type": YOU\_DECIDE\_THE\_PII\_TYPE, "text": PART\_OF\_MESSAGE\_YOU\_IDENTIFIED\_AS\_PII]\}\}}

\begin{table}[ht]
\centering
\caption{\systemname{}'s taxonomy of sensitive and personal information categories to be detected in users' prompts. We intentionally design it to be comprehensive, covering information identifiable on its own (e.g., ID), information that can be identifiable when linked with other information (e.g., demographic attributes), and personal experiences in sensitive domains (e.g., health information). The taxonomy is included in the detection prompt to instruct the LLM to generate detection results following the provided categories.}
\begin{tabular}{>{\raggedright}p{2.1cm}>{\raggedright\arraybackslash}p{5.9cm}}
\toprule
\textbf{Category} & \textbf{Definition} \\ \midrule
\textbf{NAME} & Name \\ 
\textbf{ADDRESS} & Physical address \\ 
\textbf{EMAIL} & Email address \\ 
\textbf{PHONE NUMBER} & Phone number \\ 
\textbf{ID} & Identifiers, including ID Number, passport number, SSN, driver's license, taxpayer identification number\\ 
\textbf{ONLINE IDENTITY} & IP address, username, URL, password, key \\ 
\textbf{GEO-LOCATION
} & Places and locations, such as cities, provinces, countries, international regions, or named infrastructures (bus stops, bridges, etc.).
 \\ 
\textbf{AFFILIATION
} & Names of organizations, such as public and private companies, schools, universities, public institutions, prisons, healthcare institutions, non-governmental organizations, churches, etc.
 \\ 
\textbf{DEMOGRAPHIC ATTRIBUTE
} & Demographic attributes of a person, such as native language, descent, heritage, ethnicity, nationality, religious or political group, birthmarks, ages, sexual orientation, gender and sex.
 \\ 
\textbf{TIME
} & Description of a specific date, time, or duration.
 \\ 
\textbf{HEALTH INFORMATION
} & Details concerning an individual's health status, medical conditions, treatment records, and health insurance information.
 \\ 
\textbf{FINANCIAL INFORMATION
} & Financial details such as bank account numbers, credit card numbers, investment records, salary information, and other financial statuses or activities. \\ 
\textbf{EDUCATIONAL RECORD
} & Educational background details, including academic records, transcripts, degrees, and certification.
 \\ \bottomrule

\end{tabular}
\label{tab:taxonomy}
\end{table}

\subsection{Data Minimization Support Design}

For detected sensitive entities, \systemname{} offers two data minimization methods: replacement and abstraction (\hyperref[d1]{\textbf{D1}}).
Instead of removing all identifying information (which could reduce utility), these methods substitute sensitive data with less personal equivalents. Replacement allows users to batch replace sensitive information with corresponding placeholders (\hyperref[d2]{\textbf{D2}}), useful when the detected information is not semantically essential. The abstraction method, inspired by \citet{dou2023reducing}, provides a middle ground by generalizing the detailed information while preserving context.
This is ideal when the information is relevant but overly detailed. Together, these methods give users flexible control over the granularity of data sharing (\hyperref[d3]{\textbf{D3}}).

\subsubsection{Replacement}

At the end of the detection step mentioned in \autoref{detection}, each detected item is assigned a unique placeholder label. If the user selects replacement, the tool substitutes all instances of the identified entities with their corresponding placeholders, directly applying the changes to the input box.
A tricky situation arises when one detected entity is a substring of another, such as ``15'' (age) and ``2015'' (year). To avoid incorrect replacements, \systemname{} handles this by replacing longer terms first, following a descending order of entity length. This ensures that longer entities are replaced before shorter ones, preventing misplacement of placeholders.

\paragraph{Write-back} To increase the response readability and also facilitate copying the answer (e.g., email reply), \systemname{} automatically detects the placeholders generated by the tool and writes the original entity back in the response (\hyperref[d2]{\textbf{D2}}).

\subsubsection{Abstraction} 
\systemname{} designs the abstraction mode to ensure the sanitized version not only maintains high utility but also keeps the surrounding text unchanged, making all the changes more perceptible  (\hyperref[d1]{\textbf{D1}}) and controllable  (\hyperref[d3]{\textbf{D3}}) for users. 
If the user uses abstraction, based on the current selections, the system would send a request to the LLM, asking it to rewrite the current message, abstracting all selected protected information while leaving the rest intact. The rewritten message then replaces the original text in the input box.
Below is the system prompt: \textit{Rewrite the text to abstract the protected information, and don't change other parts, directly return the text in JSON format: \{"text": REWRITE\_TEXT\}}

\subsection{Implementation}

\systemname{} is a chrome extension that consists of a frontend that displays the results and handles the user interactions, and a backend server that hosts the LLM for detection and abstraction.
According to the neural scaling laws~\cite{wei2022emergent, schaeffer2024emergent}, using larger models increases the accuracy.
This suggests a trade-off between performance and on-device deployment feasibility.
To study this trade-off in terms of user experience, data minimization efficacy, and users' trust, we implemented two versions of the tool, \modeltwo{} and \modelone{}, with the same frontend while different backend designs.
\modeltwo{}, powered by a smaller model that can run on consumer devices (e.g., Apple M-series chips with above 16GB memory) is representative of the experience of on-device data minimization support that can be achieved with current technology.
Meanwhile, \modelone{}, powered by a state-of-the-art commercial model, represents an ``upper-bound'' performance in terms of accuracy and time latency.
We use \modelone{} as a reference point to assess the performance of \modeltwo{} objectively and subjectively.
We detail the implementation of the two versions below.

\subsubsection{\modeltwo{}}
We serve the Llama3-8b (4-bit quantized) model using the Ollama framework\footnote{\url{https://ollama.com}}, which is a lightweight, extensible framework for building and running language models on the local machine.
The model's temperature was set to 0 (greedy sampling) to ensure deterministic detection and abstraction results.
The K/V prompt cache has been enabled to avoid processing the system prompt repeatedly, which helps save time.

\paragraph{Input segmentation}
When a user submits a message, Rescriber processes the input by breaking down larger text into smaller chunks. This approach addresses the known issue of smaller LLMs having difficulty with processing long prompts and sometimes forgetting the instructions presented at the beginning. Additionally, chunking allows for potential parallel processing as a future optimization direction to further accelerate response times.

\paragraph{Streaming mode}
We made a key observation about the time performance that generating output tokens was more time-consuming than processing input tokens due to the inherently sequential nature of the output generation process.
Hence, we introduced the streaming mode, which continuously parsing a stream of output tokens from the model, and immediately display each newly detected entity to users rather than waiting for the entire input to be processed.
Streaming reduces the delay for tasks like PII detection and abstraction. Since reviewing detected items or abstracted terms often requires time and attention, this makes the system more responsive by parallelizing response generation and user evaluation of the results.

\subsubsection{\modelone{}} We used the OpenAI API endpoints to implement this version of the system, with the temperature set to 0.
However, GPT-4o model is known to be non-deterministic even when the $temperature = 0$.
This is caused by its use of Sparse MoE.
This is an inherent limitation of the model's architecture that turns out to affect users' preferences as we later discovered from the user studies. 

\paragraph{Clustering}

Despite effective PII detection, we identified issues related to clustering variants of the same entity. For example, in pilot testing, the extension successfully masked ``[first name] [last name]'' but failed to replace ``[first name], [last name]'', resulting in a suboptimal user experience. This highlights that partial masking may create a false sense of security or heighten discomfort if users become aware of incomplete data minimization.
To address this, we further built in a same-entity clustering component in the \modeloneplain{}.
The \modeltwoplain{}, due to its smaller size and limited capabilities, skips clustering because it is resource-intensive and time-consuming.

\subsubsection{Performance benchmarking}
\label{sec:implementation-eval}
We conducted a small experiment to benchmark the accuracy and time performance of the two models to establish an objective comparison of their performance.
We also selected Microsoft Presidio\footnote{\url{https://microsoft.github.io/presidio/}}, a well-known PII detection SDK, as another baseline to represent the state-of-the-art accuracy of traditional NLP methods for PII detection (e.g., regular expressions, named entity recognition).
This serves as a reference to users' subjective preferences and data minimization actions enabled by the extension in our user studies.
Our evaluation dataset comprises 240 randomly selected samples from Ai4Privacy’s world’s largest open dataset for privacy ~\cite{ai4privacy_pii_masking_200k}, which encompasses 229 discussion subjects and use cases.
These samples were processed by aligning the 54 PII classes from the dataset with the categories in our taxonomy (\autoref{tab:taxonomy}).
To ensure accuracy, two authors independently reviewed and verified the privacy information labels and the corresponding text.
Their consensus was used as the ground truth.
\autoref{tab:precisionandrecall} shows that both \modelone{} and \modeltwo{} outperforms Presidio.
Both models achieved high precision, while the recall of \modelone{} was higher than \modeltwo{}.
For time latency, we benchmarked the small models (Presidio and \modeltwo{}) with four device configurations.
\autoref{tab:response_times} presents the time latency results with an M2 chip, 24GB memory setting, and the full results can be found in \autoref{sec:full-time-benchmark-result}.
As we expected, the streaming mode greatly reduces the time to display a detection result to users from an average of 3.40s to 2.01s (69\% reduction).
Notably, with an M1 Max Macbook Pro (64GB memory), the average latency of generating the full response using \modeltwo{} (1.09s) was similar to Presidio (0.84s) with the same device settings and faster than GPT-4o (1.41s).
The source code and instructions for reproducing the study results are included as supplemental materials.

\begin{table}[h!]
\centering
\caption{Comparison of Precision and Recall (with Standard Deviations) among Presidio, \modeloneplain{} and \modeltwoplain{}.}
\resizebox{\columnwidth}{!}{%
  \begin{tabular}{lcccc}
    \toprule
    \textbf{Model} & \textbf{Precision} & \textbf{Recall} & \textbf{Precision SD} & \textbf{Recall SD} \\
    \midrule
    \modelone{}  & 0.94 & 0.88 & 0.15 & 0.21 \\
    \modeltwo{}  & 0.94 & 0.63 & 0.16 & 0.28 \\
    Presidio    & 0.69 & 0.68 & 0.27 & 0.31 \\
    \bottomrule
  \end{tabular}%
}
\label{tab:precisionandrecall}
\end{table}

\begin{table}[h!]
\centering
\caption{Comparison of Response Times Among Presidio, \modeloneplain{} and \modeltwoplain{}. On-device models are benchmarked with different device configurations. This table only shows the results with the M2 chip 24GB memory setting. Full results can be found in the appendix (\autoref{tab:full_response_time_table}).}
\resizebox{\columnwidth}{!}{%
\begin{tabular}{lcccccc}
\toprule
\textbf{Model} & \textbf{Min (s)} & \textbf{Max (s)} & \textbf{Mean (s)} & \textbf{SD (s)} \\
\midrule
\modelone{}  & 0.53    & 3.98    & 1.41    & 0.65              \\
\modeltwo{} (first detection)  &  1.34  &  4.52  &  2.01   &      0.48     \\
\modeltwo{} (full detection) &  1.62  &  7.93  &  3.40   &      1.29     \\
Presidio &    0.60  & 1.98   &  0.73   &    0.19       \\
\bottomrule
\end{tabular}
}
\label{tab:response_times}
\end{table}

\subsubsection{On-Device Feasibility Test}
\label{sec:on-device-feasibility-test}

To evaluate the performance and feasibility of our on-device \modeltwo{} (with streaming mode enabled), we conducted IRB-approved remote user studies with eight participants from diverse backgrounds (4 non-tech, 2 tech, and 2 mixed). Participants were required to have a device with an M-chip and at least 16GB of memory (see \autoref{Ondevice} for participant and device details). The study aimed to assess the usability and real-world applicability of deploying a small on-device LLM for average users.
Participants were given an installation manual and asked to install the extension on their devices. They then tested it using a hypothetical email-writing task. To measure perceived performance, we used the UMUX-LITE scale~\cite{lewis2013umux} and included a question about satisfaction with detection speed. All eight participants successfully installed and used \modeltwo{}. \autoref{tab:on-device-result} shows the median evaluation scores. The results indicate that our system operates with a user-accepted speed of detection and usability.

\begin{table}[h!]
\centering
\caption{On-Device Usability Result. The median user ratings (1=Strongly disagree; 7=Strongly agree) are shown for each item.}
\resizebox{\columnwidth}{!}{%
\begin{tabular}{l c}
\toprule
\textbf{Item}               & \textbf{Agreement} \\ \midrule
Capabilities meet requirements   & 6 \\ 
Not a frustrating experience               & 6  \\ 
Easy to use         & 5   \\ 
Not spend much time correcting things with this system                    & 6 \\ 
Comfortable with the detection time                    & 6.5 \\ \bottomrule
\end{tabular}
}
\label{tab:on-device-result}
\end{table}

\section{Methodology}
We conducted twelve semi-structured online interviews with participants from the U.S to investigate our research questions around how users perceive and interact with a tool designed to facilitate user-led data minimization and how AI models with varying levels of accuracy influence the subjective perceptions and objective effectiveness.
The study was approved by our institution's IRB.
Note that we chose to set up the Ollama server on a Google Cloud Platform virtual machine for the interviews to relax the requirements of our participants' devices and achieve consistent time performance during the studies.
Our performance benchmarking (\autoref{sec:implementation-eval}) and on-device feasibility user testing (\autoref{sec:on-device-feasibility-test}) have validated the practicality of the on-device version of Rescriber.
The virtual machine features 4 vCPUs and an NVIDIA L4 GPU (24GB VRAM).
The streaming mode was not enabled during these interviews\footnote{The streaming mode was introduced after conducting these interviews. Since the interviews' goals and findings focused more on the impact of the interaction paradigm and the \systemname{} suggestion accuracies, not enabling the streaming mode does not affect our findings.}.

\subsection{Participants}

Participants were recruited through Prolific, an online research recruiting website. The recruitment message for the screening survey did not mention privacy. We employed screening criteria that required participants to have some experience using ChatGPT and privacy concerns related to its use. This ensured that participants had related experiences and could provide real-world prompts for the testing. Of the \totalsamplesize{} participants who completed the initial screening survey, \qualifiedsamplesize{} expressed privacy concerns and willingness to participate in a 1-hour follow-up remote interview study.
Participants were compensated \$20 USD. We randomly selected 28 of these participants and distributed a pre-study survey for interview signup and preparation. \samplesize{} participants successfully completed the interviews.

According to their self-reported data, all participants were fluent English speakers. The age distribution was as follows: 3 were aged 18–24, 6 were 25–34, 1 was 35–50, and 2 were over 50.
Among those interviewed, 6 used ChatGPT several times a day, 2 used it once a day, 2 used it several times a week, and 2 used it less than once a week. All participants had concerns about exposing personal information (e.g., demographic information, health details, educational records) to the model, and \tookminimizationactionsize{} self-reported to have tried to minimize unnecessary information sharing with ChatGPT.

\subsection{Study Design}

Our study design involves four sessions, consisting of having two conversations with GPT using our extension powered by two different back-end models. 
The two conversations included one participant's shared example, while the other one used a hypothetical scenario prepared by the research team, addressing common use scenarios inspired by the ShareGPT dataset~\cite{sharegpt52k}.
Each participant was assigned one of three hypothetical scenarios. 
The three scenarios include reading and responding to an email, performing data analysis on a dataset, and writing a thank-you letter to a therapist post-treatment (see \autoref{sec:hypothetical-examples} for details).

The study followed a counterbalanced design to reduce the order effects.
Participants were divided into groups, with half testing the \modeloneplain{} (referred to as Model 1) first and then the \modeltwoplain{} (referred to as Model 2), and the order of using their own example was also alternated.
They always started by testing one example (their own example or the hypothetical example) on two models sequentially, and then switched to testing another example on two models in the same order.
This design allowed the participant to make direct comparison between the experiences with the two models on the same example.

\subsection{Interview Procedure}

During the interviews, participants received an information sheet outlining their rights and shared only what they were comfortable with. We also obtained their consent to record the session for note-taking and post-interview data analysis.
Once recording commenced, we asked them to introduce the typical ChatGPT use cases, their privacy concerns and data minimization experiences (if any).
Participants were asked to select a real-world conversation they had with GPT involving personally identifiable information, personal experiences, or other sensitive details they felt comfortable sharing with the research team.
They were instructed to prepare this conversation when signing up but could also review their chat history if unprepared.
The moderator requested a brief explanation for their selection, confirming that both the context of the message and the task for ChatGPT were understood and the shared personal information was consented to share.

After reviewing the prepared message, participants watched a 3-minute demo video (included in the supplementary materials) demonstrating the extension and its functionalities. They were then asked to describe their understanding of the system's capabilities to ensure comprehension. We introduced the study's four sessions, and instructed participants to think aloud and interact with ChatGPT as they would in a realistic scenario, including asking follow-up questions to fulfill their requests.
To streamline the procedure and avoid accidentally recording users' chat history, participants controlled the experimenter's pre-configured laptop and used the experimenter's ChatGPT account.
To avoid bias, the two models were referred to as Model 1 and Model 2, and we debriefed the details of the two models at the end of the interview.

Participants were asked to explain their choices during testing and provide feedback on their satisfaction with ChatGPT's post-sanitization responses, perceived privacy protection, and reduction of unnecessary information sharing with ChatGPT.
After each session, participants were asked to complete a follow-up survey, rating agreement with the following statements on 5-point Likert scale:

\begin{itemize} \label{questionnaire}
\item The extension reduces the disclosure of unnecessary information to ChatGPT.
\item The extension reduces the disclosure of my personal information to ChatGPT.
\item I have fewer privacy concerns when talking to ChatGPT using the extension than without using the extension.
\item I would use this extension, assuming it is well-designed in terms of usability.
\end{itemize}

After finishing testing one example with two models, participants were asked to explain any differences in their ratings across the two sessions, helping us identify which features of the models influenced their trust in the tool.

\subsection{Qualitative Analysis Method}

We conducted a bottom-up qualitative coding analysis using affinity diagramming~\cite{beyer1999contextual}. All interview transcripts were imported into Figjam, an online collaborative whiteboard. In the first stage, four researchers reviewed recordings of four interviews and wrote interpretive memos on notable excerpts relevant to our research questions. The researchers met twice a week to collectively conduct affinity diagramming and develop clusters of memos. At the end of this stage, the researchers refined the clusters and created names for each to form an initial codebook.

In the next stage, the remaining eight interviews were evenly assigned to the researchers, with each interview reviewed by two researchers. They used the initial codebook to code and suggest changes based on emerging patterns. Regular meetings were held to refine the codebook together. The final codebook can be found in \autoref{sec:codebook}.

\subsection{Methodological Limitations}
There are several methodological limitations that need to be considered when interpreting the results.
First, although our participants shared many real-world use cases that could benefit from our tool, including legal inquiries about family matters, planning travel or medical arrangements, creating 401(k) distribution plans, and drafting emails for seeking lab opportunities, many felt uncomfortable sharing a prompt they used in these cases in a recorded session, and chose to share a different prompt which contained less personal information.
This could influence their perceptions of the usefulness of the tool.
It is noteworthy that our three hypothetical test cases covered several use cases participants mentioned, such as proofreading (E1) and asking GPT to analyze a dataset (E2), which helps alleviate this issue.
Another limitation is due to the nature of the controlled usability study, in which participants interacted with the tool under the observation of the researcher.
The participants may be affected by the Hawthorne effect and tend to speak in favor of our tools.
To address this issue, we referred to the two versions of the system as Model 1 and Model 2, without revealing the main evaluation target, and encouraged participants to voice negative comments.

\section{Quantitative Results}
\label{sec:quant-findings}

All 12 participants completed every task. We examined key metrics such as the number of attempts per participant, satisfaction, and sanitization efforts (replacements and abstractions) to compare the models' performance. There was one error in the counterbalancing: for P6, the interview was supposed to start with \systemname{}-Llama3-8B followed by \systemname{}-GPT-4o, but the moderator mistakenly began with \systemname{}-GPT-4o.

\subsection{Data Minimization Efficacy}
\label{sec:quant-findings-behavioral}
We allowed the participants to experiment with each example multiple times.
For the three hypothetical examples (E1-E3), each participant made 1.1 attempts per example on average.
For user's own example (E4), the average number of attempts was 1.4, showing that users explored different ways to sanitize their prompt using \systemname{} a bit more than in the three examples provided by us.

\begin{table}[h!]
\centering
\caption{Overview of the data minimization efficacy evaluation. The column ``Satisfaction'' indicates the number of attempts that resulted in a satisfactory response. Among all attempts,  84\% led to a satisfactory response. We then show the average number of replacement and abstraction performed in the \textit{satisfactory attempts}, which indicate the successful reduction of unnecesary disclosure.}
\resizebox{\columnwidth}{!}{%
\begin{tabular}{c c c c c}
\toprule
\textbf{Example} & \textbf{Satisfaction} & \textbf{Average replacement} & \textbf{Average abstraction}  \\ \midrule
E1 & 8 out of 9 & 2.3 & 1.4  \\ 
E2 & 7 out of 8 & 26.4 & 3.6  \\ 
E3 & 9 out of 9 & 1.7 & 1.1\\ 
E4 & 25 out of 32 & 1.0 & 0.3\\
\bottomrule
\end{tabular}
}
\label{tab:example-sanitization}
\end{table}

We first evaluate \systemname{}'s efficacy in data minimization, measured by the amount of reduced disclosure in attempts that resulted in a satisfactory response (\textbf{RQ1}).
The results are summarized in \autoref{tab:example-sanitization}. Among all attempts, 84\% led to a satisfactory response. Users reduced more unnecessary disclosure in the examples we provided (E1-replying email: 3.7, E2-data analytics: 30.0, E3-writing a letter: 2.8) than in their own example (E4: 1.3). Note that E4 is a collection of unique examples from each participant.
The lower reduction in E4 may be due to participants being less comfortable sharing personal cases in a recorded session.
These results indicate that \systemname{} successfully helped users reduce unnecessary disclosure, with the amount depending on context.
We found that the average number of replacements consistently surpassed that of abstractions, which suggest that users found replacements more suitable in more situations.
Furthermore, comparing the two models, \modelone{} achieved substantially more reduction of unnecessary disclosure (8.0 in total, 7.4 replacements, 0.6 abstractions) than \modeltwo{} (4.1 in total, 2.6 replacements, 1.5 abstractions).

The three most frequently replaced or abstracted categories are NAME, GEOLOCATION and TIME.
We see that both models can use categories that are outside of the defined taxonomy (\modelone{}: INSTITUTION, URL; \modeltwo{}: AGE, USERNAME, WEIGHT, HEIGHT, GENDER, NATIONALITY).
\modeltwo{} seems to have an edge in handling abstraction and replacement of more concrete entities (e.g., GENDER) rather than using the pre-defined category (e.g., DEMOGRAPHIC ATTRIBUTE).

\subsection{Response Utility}
\label{sec:response-utility}
For the data minimization attempts that led to satisfactory responses, we further measured the utility preservation by comparing them with the responses generated with the original prompt.
LLM-as-a-judge~\cite{10.5555/3666122.3668142} has become a popular approach for assessing a pair of LLM responses to the same question, which offers the benefits of scalability and explainability and achieves high consistency with human ratings.
We adopted this approach and referred to the prompt design of \citet{10.5555/3666122.3668142} in our experiments.
GPT-4o was used both for response generation and as the judge.
Note that the self-enhancement bias was not an issue here because both responses rated by GPT-4o were generated by itself.

The system prompt instructs GPT-4o to grade the two responses based on a 5-point Likert scale with specific considerations on format and content, generating brief reasoning for each decision. The full system prompt is referred to in \autoref{sec:response-comparison-prompt}. To guarantee unbiased comparison results, we generated ten responses per prompt (original and masked) and randomized their positions to eliminate randomness effects and position bias.

\begin{table}[h!]
\centering
\caption{Average similarity scores of all participants' satisfactory attempts for each example (E1, E2, E3, E4) and redacted using \modeloneplain{} and \modeltwoplain{}. For each attempt, GPT-4o compared responses generated from the original input message (before redaction) with responses generated from the redacted message (after redaction using \systemname{}). Similarity ratings were based on a 1–5 Likert scale: 1 and 5 indicated a strong preference for the original or replaced-back response, respectively; 2 and 4 indicated a slight preference, and 3 indicated comparable quality. Evaluations covered two dimensions: format and content.}
\begin{tabular}{|c|cc|cc|}
\hline
 & \multicolumn{2}{c|}{\modelone{}} & \multicolumn{2}{c|}{\modeltwo{}} \\ 
\textbf{Example}  & \textbf{Content}        & \textbf{Format}        & \textbf{Content}             & \textbf{Format}             \\ \hline
E1               & 2.8                & 2.7               & 2.9                     & 2.7                    \\ \hline
E2               & 2.4                & 3.1               & 2.4                     & 3.0                    \\ \hline
E3               & 2.9                & 3.0               & 2.8                     & 2.9                    \\ \hline
E4               & 2.9                & 3.0               & 2.7                     & 3.0                    \\ \hline
\end{tabular}
\label{tab:utility_comparison}
\end{table}

Table~\ref{tab:utility_comparison} presents the average similarity scores of all participants' satisfactory attempts for each example (E1, E2, E3, E4).
Both versions of Rescriber achieved similar results.
We observe that E1, E3, and E4 scores near 3, which suggests that redaction minimally affected response utility. E2's lower content scores reflect the challenges of GPT-4o in interpreting a table with a lot of placeholders in data analysis tasks. 
All four examples had a format score close to 3, indicating a good preservation of the responses' format.
E1 had a slightly lower format score (2.7).
This is perhaps because GPT-4o tends to propagate placeholders when encountering them in the input text. For instance, when redacted messages include structured placeholders like \texttt{[NAME1]}, the generated response sometimes introduces unrelated placeholders such as \texttt{[Your Full Name]}.
Examples can be found in \autoref{sec:response-comparison}. We discuss the limitations with current LLMs in processing redacted messages and potential improvements in \autoref{sec:discussion-small-model-based-control}.

\subsection{Subjective Preferences}
\label{sec:quant-findings-subjective}
\autoref{tab:subjective-results} summarizes the analysis results of users' subjective ratings of the two versions of \systemname{} after use.
The four questions focused on the reduction of unnecessary disclosure, perceived reduction, privacy concerns, and intention to use, rated on a 5-point Likert scale (higher is better).
The median ratings were identical for reduction of disclosure and intention to use, with \modeltwo{} slightly lower for the other questions.
We further performed the Wilcoxon signed-rank test to compare the subjective ratings between the two models and found no statistically significant difference between the two models for all four questions.
This suggests that participants viewed both models as relatively equivalent in terms of data minimization, privacy concern mitigation, and the adoption intention.

\begin{table*}[h!]
\centering
\caption{Comparison of models based on Wilcoxon test results (1=Strongly disagree; 5=Strongly agree)}
\begin{tabular}{l c c c c}
\toprule
\textbf{Question}               & \textbf{\modelone{}} & \textbf{\modeltwo{}} & \textbf{Statistic} & \textbf{P-value} \\ \midrule
Reduce the disclosure of unnecessary information   & 5.0              & 5.0              & 29.5               & 0.4406           \\ 
Reduce the disclosure of personal information               & 5.0              & 4.5              & 24.5               & 0.1294           \\ 
Fewer privacy concerns using the extension         & 5.0              & 4.0              & 21.5               & 0.1494           \\ 
Would like to use the extension                    & 5.0              & 5.0              & 11.0               & 0.6089           \\ \bottomrule
\end{tabular}
\label{tab:subjective-results}
\end{table*}

\section{Qualitative Results}

We summarize the qualitative analysis results, including users' privacy concerns with ChatGPT, their natural data minimization strategies and the limitations (\textbf{RQ1}); factors that affect users' trust in \systemname{}'s ability to enhance privacy (\textbf{RQ2}); factors that affect users' intention to use \systemname{} (\textbf{RQ2}); and the data minimization strategies in reaction to the tool, as well as the educational effect and learning curve (\textbf{RQ3}).

\subsection{Privacy Concerns, Yet Why Still Using ChatGPT?}
\label{sec:qual-findings-concerns}

We start by examining privacy concerns related to ChatGPT, how users address them, and why they continue using it.
These factors influence ChatGPT users' decision on data disclosure and sanitization, reinforcing the need for user-driven data minimization. 

\subsubsection{Concerns with identifiability (All but P4, P7, P10)}
\label{sec:qual-findings-concerns-identifiability}
A lot of participants expressed concerns about various data types that could potentially reveal their identities when sharing information with ChatGPT, including direct personal identifiers, such as names (P1, P2, P3, P6, P8, P9, P12) and contact information (P1, P2, P8, P11).
P3 emphasized the importance of names when testing with the hypothetical email reply example (E1), stating
\blockquote{the most important thing to abstract is the names.}

Concerns about identifiability also extended beyond direct identifiers to demographic details like gender, raising fears of linkage attacks~\cite{vatsalan2013taxonomy} and broader profiles emerging from multiple data points.
P5 stated,
\blockquote{when they [my height and weight input] are all taken together, ...it paints a different picture.}
Prolonged interactions increased the risk, as mentioned by P9,
\blockquote{throughout all the questions I asked ChatGPT..., [ChatGPT] probably can paint the picture of who I am.}
Possessive terms also raise concerns. P3 provided an example: \blockquote{Do Indian people ...? What series tend to be the favorite of people like me?} While the term ``Indian people'' was not seen as particularly concerning, the possessive phrase ``people like me'' created an identifiable connection.

\subsubsection{Concerns due to lack of transparency (P3, P5, P7)}
Many participants linked privacy concerns to fear of the unknown, due to a lack of transparency about how their data is handled by the LLM-based model.
P5 illustrated this apprehension: \blockquote{It’s like when a credit card company asks for your social security number: I don't know where that's going... same way my privacy concerns are heightened here.}
P7 echoed this sentiment, noting, 
\blockquote{it [ChatGPT] is a system that we really don't know the inner workings of, so you don't know what risk you're running when you share information with it.}

\subsubsection{Harms (P4, P7, P9, P10)}
Concerns ranged from general risks to specific harms. Some participants (P7, P10) were unclear about specific threats but believed harms were likely. P10 voiced a broad worry: \blockquote{I fear that sharing my real information... could end up in the wrong hands.}
Others, like P4, worried about specific harms such as financial risks that could arise from sharing bank account information, suggesting it could lead to fraud: 
\blockquote{a scammer...try to brute force into an account that I might have at this particular bank... could be problematic.}
Legal risks also concerned participants; P4 mentioned
how sensitive legal research could raise red flags with authorities.

\subsubsection{Natural data minimization strategies} 
Our interviews revealed five primary data minimization strategies that participants naturally employed without the aid of our tool: 

\textbf{Halting conversations (P4).}
    People chose to terminate conversations to avoid further sharing. \blockquote{I didn't necessarily change the prompts that I've already set. But I did stop the conversation,} suggesting a strategy to avoid escalating personal exposure once a conversation has begun.

\textbf{Avoiding records (P3, P4, P9).}
    Many participants were conscious of leaving a permanent record with ChatGPT, particularly due to concerns about its learnability and memorization as an LLM-based system. P9 explained they toggle off the memory, stating, \blockquote[]{I'm traveling... I have a plant at home... whose leaves turn yellow... All those things were documented in the memory, which was certainly not something I want.}
    P4 would delete the chats to erase the records, underscoring a desire to avoid a traceable digital footprint. 

\textbf{Omitting details (P1, P8, P11).} 
    Several participants intentionally omitted key personal details, such as names or financial information. P11 explained, \blockquote[]{I just didn't give my name.} Similarly, P1 avoided including financial details for business purposes, \blockquote[]{If I'm ever using it to build my business, I don't put in... personal information.} These examples illustrate a deliberate effort to limit the sensitivity of data shared.

\textbf{Faking data (P4, P5, P9).}  
    Some participants provided false details to mask their identity or situation, often altering the subject of the stories or inquiries to avoid linking the information directly to themselves. For example, P5 and P9 shared concern about making data traceable, prompting them to change names. 
    Additionally, participants obscured details even when they were not personally involved.
    P4 shared an example: \blockquote[]{I have an employee who had a baby... I chose to say friend instead of employee... it seems sensitive.} 

\textbf{Generalizing information (P2, P8).}
    Another common strategy involved generalizing queries to avoid sharing precise or identifiable details. For example, P8 noted that when submitting files for review, they would alter or summarize the content, \blockquote[]{I try to edit it a little bit or do summary... without super important, more confidential [information].}

Participants acknowledged, however, that even with these strategies, they couldn’t fully control their data once shared. P9 noted that though they attempted to avoid certain actions, the system's memory might still retain their information.
Similarly, P4 articulated a fear of losing control, believing the data may never truly be erased once shared.

\subsubsection{Continuous usage and trade-off (P3, P4, P9)}
Despite concerns, all participants continued to use ChatGPT due to its convenience. As P4 mentioned, they were willing to \blockquote{risk the privacy} to \blockquote{save the time}; while P9, despite worries about exposing financial data, used GPT for financial advice.
P3 captured the ``inevitable'' trade-off, stating, \blockquote[P3]{I think it's kind of a dilemma right? Because I do try to avoid giving more information than necessary, but it feels... difficult to get what I want..., without providing... information inevitably.}.
These struggles highlight a clear demand for privacy-preserving data minimization support in LLM-based conversational agents.

\subsection{Factors for Privacy Trust}
\label{sec:qual-findings-trust}
We identified six factors impacting user privacy trust over interaction with ChatGPT and \systemname{} to evaluate and inform the design of a trusted, user-led data minimization tool.

\subsubsection{Transparency and On-Device Needs of the Extension (P4, P7)}
A key factor influencing user privacy trust is the need for transparency and an on-device implementation of the data minimization tool, which aligns with our goal of leveraging smaller models hosted locally (e.g., \modeltwo{}). 
Participants stressed the importance of understanding how their data is managed. \blockquote[P7]{that's just a matter of transparency... you never know that they're not [collecting information].} Participants expected local processing to avoid replicating their concerns with ChatGPT.
P4 emphasized, \blockquote[P4]{I assume that the extension is running itself locally and is not storing any of the data...}, reflecting users' need for control over their data. 

\subsubsection{Consistent Detection Results (P4, P9, P11)}
A critical factor affecting user trust in \systemname{} was the consistency of detection results.
Participants expressed frustration when the similar inputs produced different feedback, which undermined their confidence in the tool's reliability.
P11 noted confusion when one model detected fewer privacy risks than the other, whereas P9 observed that \systemname{} missed sensitive information when given partial inputs, while caught them when given the full messages.
Similarly, P4 expressed skepticism when \modelone{} flagged sensitive content during a second run not the first: \blockquote[P4]{it loses... credibility... making me skeptical of robustness of the tool.} This lack of consistency, sometimes due to the non-deterministic nature of the model, led participants to question the effectiveness of \systemname{}'s capabilities.

\subsubsection{Consistent Abstracting Results (P5, P10)}

Users' confidence in \systemname{} is closely tied to the consistency of its abstraction results. Inconsistent abstraction can lead to doubts about the system’s ability to protect sensitive data. P5 expressed concern when \modelone{} displayed inconsistency in abstracting ``KCMO'' also due to its non-deterministic nature, first changing it to ``major city in Midwest'' and later to ``my city,'' prompting P5 to question the integrity of the responses.
In contrast, P10 reported satisfaction with \modeltwo{} for its consistent abstraction, noting, \blockquote{I'm satisfied because the same information is again over here}.

\subsubsection{Completeness of Detection (P2, P4, P5, P6, P8)}
\label{sec:qual-findings-trust-complete-detection}
A key factor influencing user trust in \systemname{} is the completeness of its detection and sanitization. Participants felt more confident and protected when the system detected a broader range of sensitive data. P5 highlighted the tool’s ability to catch overlooked details, \blockquote{it identified things that I didn't even pick up could be sensitive}.

Participants often compared models to assess detection. P5 preferred \modeltwo{} which detected more identifiers like height and weight, while P2 found \modelone{} more thorough, \blockquote{we picked up way more on this one... so this one is doing much better.}
These examples are related to the advantages of the two models in different aspects.
\modelone{} flagged more entities overall, as echoed by our quantitative findings (\autoref{sec:quant-findings-behavioral}); and \modeltwo{} excelled at detecting entities outside of the specified taxonomy, possibly due to its weaker instruction-following capabilities which made it more likely to not be constrained by the taxonomy.

Participants' faith in the system diminished with incomplete detection. P6 expressed doubt when they 
\blockquote{don't necessarily know that it caught all the personal information.}
The consequence could be catastrophic.
P4 encountered a case where the model failed to detect a piece of sensitive information he expected it to catch, completely disrupting his trust. He remarked, \blockquote[]{It loses all of its reputability with me right away.} 
These examples reinforce the importance of thorough detection for fostering user trust in privacy-preserving tools like \systemname{}.

\subsubsection{Completeness of Sanitization (P3, P4, P6, P7)}
\label{sec:qual-findings-trust-complete-sanitization}
Participants emphasized the importance of providing sanitization features to ensure complete data minimization.
\systemname{}'s ability to aggregate all instances of the same data type was praised by P4, stating, \blockquote[]{I really enjoyed that everything was sorted properly... I'm not missing one because they're all there.}
The ``select all'' option was also valued for simplifying complete privacy protection, as put by P7, \blockquote[]{that would be much easier if you would do that in one click.}
Participants sometimes even expressed a desire for more assertive privacy-preserving mechanisms. P6 remarked on the system's limitations in enforcing data minimization: \blockquote[P6]{it helps me reduce it, but it doesn't necessarily force me to do that,} reflecting a wish for a more proactive approach to completely redact sensitive data.

\subsubsection{Control over Sanitization (P4, P5, P6, P7)}
\label{sec:qual-findings-trust-control}
Another factor in building user trust in \systemname{} is giving participants control over the sanitization process. Participants reported greater trust and satisfiction given the autonomy to decide which data to mask.
\blockquote[P6]{I have more control over the information I'm sharing.}.
P5 further highlighted the empowerment from this control, explaining, stating, \blockquote[P5]{it made me aware of those different areas and gave me... the final choice.} This sense of control allowed users to feel in charge of their privacy decisions.

\subsection{Factors for Adoption}
We identified key factors influencing user adoption of \systemname{} for privacy-preserving data minimization practices.

\subsubsection{Balancing Privacy with Task Effectiveness (All but P4, P7, P11)}
\label{sec:privacy-trade-off-balance}
\systemname{}'s ability to generalize sensitive information without compromising task effectiveness influenced user adoption. Participants
were satisfied when abstractions preserved contexts. For instance, 
P3 approved of abstracting ``Yonkers, NY'' to ``a city in NY,'' as this generalization still allowed them to find recommended places within the region. P10 valued abstracting ``John Doe'' with ``a traveler'' in E2, which kept the travel data analysis task intact.
Nevertheless, when abstraction distorted the original intent, participants 
were less likely to adopt the system’s solutions. P6 rejected the abstraction of ``downtown Troy'' to ``State Park,'' explaining that the context had shifted too far from their original intent.

Participants were not always able to predict the information that optimized privacy-utility trade-offs. P2 removed some health information from her prompt as suggested by \systemname{}, expecting less thorough feedback.
However, after comparing the new response based on the redacted prompt with the original, she found them very similar, and said, \blockquote{I think I'm satisfied... it doesn't seem to be as detrimental as I thought.}

\subsubsection{Convenience (P3, P7, P8, P10, P11)}
Overall, participants valued the convenience provided by \systemname{}, recognizing its ability to simplify complex privacy-preserving tasks, which drives user adoption.
P7 emphasized the importance of simplicity, noting it, \blockquote{simplify the whole process of using something that's already supposed to be easy.}
Another key convenience factor was the available features that streamline data sanitization:
\label{sec:qual-findings-convenience-features}
\begin{itemize}
    \item \textbf{Select all:} P8 appreciated the ``select all'' feature, saying, \blockquote{convenience and just helping you get stuff done at a faster rate.} 
    \item \textbf{Replace back:} Participants noted that replacing sanitized terms back into the conversation improved readability. P3 remarked on its usefulness for comprehension, stating, \blockquote{really useful for keeping comprehension up, because if we just gave the default names, you'll get really confusing.} 

\end{itemize}

\subsubsection{The tool should respect people's habit of using ChatGPT (P4, P6, P8, P11)}
Another factor influencing user adoption of \systemname{} is how the tool affects their regular interaction patterns with ChatGPT. Participants expressed a reluctance to change their established workflow, which prioritizes the speed and fluidity of exchanges with the chatbot. For instance, P4 described their typical behavior when using ChatGPT as a back-and-forth interaction without much interruption, \blockquote[P4]{It's a shame that I would have to change the way that I operate with ChatGPT, because my instinct is to treat it like a chat bot, just input...type [enter to send].. input.. type [enter to send]... input... 
} This highlights a friction between the participants' habitual use of ChatGPT and the interruption caused by privacy protection measures.

\subsubsection{Having Alternative Solutions as a Benefit (P3, P5)}
\label{sec:qual-findings-alternative-solutions}
One final key adoption factor was the availability of actionable solutions. Participants appreciated that \systemname{} not only identify sensitive information but also offer practical alternatives to sanitize it. P3 highlighted the challenge of manually abstracting sensitive information: \blockquote[]{having something that can do it for you is really useful.} P5 similarly valued the tools providing actionable solutions, rather than just flagging sensitive content: \blockquote[]{provides solutions instead of just marking a word.} 

\subsection{Data Minimization Strategies in Reaction to the Tools}

We uncovered nuanced behaviors and decision-making processes that participants adopted to minimize the data they shared with ChatGPT during testing. Their decisions were influenced by the task context and sensitivity of the data.

\subsubsection{What to sanitize: balancing sensitivity and task relevance}
\label{sec:qual-findings-tool-sanitization-strategies-datatype}

Participants tailored data sanitization based on whether masking certain information would impact the quality of the response. They aimed to obscure personal or sensitive data without compromising the model's output, but opinions differed on what constituted sensitive information.

\textbf{Task relevance (P6, P11).}  
Some participants removed unnecessary details that were perceived to not affect the task, like P6 who noted that details about why someone visited a treatment center weren’t needed for writing a letter.
Conversely, P11, argued that certain details could not be obscured without negatively impacting the conversation’s purpose, particularly, in medical contexts where specifics are crucial. 

\textbf{Removing partial information as a compromise (P2, P4, P5, P6, P11).}  
Many participants preferred to obscure only part of the identifiable data, believing that hiding one or two key details was enough to preserve privacy and get useful feedback. P5 felt that altering height or weight alone could suffice for maintaining privacy while not affecting the task, explaining that, \blockquote[P5]{changing one of them... would be sufficient.}

\textbf{Individual differences (P2, P5, P7).}  
We also observed individual differences on data minimization strategies over a same case, E3, about writing a thank-you note to the hypothetical user's therapist. For example, P5 was comfortable obscuring some health details when writing a thank-you note to a therapist, while P2, worried that over-sanitizing health-related information would lead to overly generic or less useful responses: \blockquote[P2]{If we're keeping this stuff out of GPT, it's not going to be able to provide as thorough of an answer.}
P7 emphasized that for domain-specific tasks (e.g., legal questions) people may have different levels of expectations of the specificity of the answer based on their domain knowledge, explaining, \blockquote{it does help protect your privacy if you don't know much about the law and you're just getting a baseline...it depends on the knowledge level of the person using it.}

\textbf{Name data as a high-priority concern (P1, P3, P4).}  
Across multiple participants
, names emerged as a particularly sensitive data type, with many users prefer to remove or anonymize them, echoing the quantitative analysis result which showed name as the most frequently redacted or abstracted data type (\autoref{sec:quant-findings-behavioral}).
P3 noted that names were non-essential for many cases and should be removed first, which reflects a common concern across participants.

\subsubsection{Rationales for choosing replacement or abstraction}
\label{sec:qual-findings-tool-sanitization-strategies-actiontype}

Participants adopted different strategies to choose between replacing data with placeholders or abstracting the information, depending on the context and their privacy concerns.

\textbf{Abstraction for retaining context (P3, P5, P7, P8).}  
Abstraction was favored for conveying general contextual information without revealing specifics. For example, P3 opted for abstracting numbers to maintain content clarity without divulging precise information, stating that abstraction helps \blockquote[]{get an idea without providing specific numbers.}

\textbf{Replacement for non-essential data (P3, P12).}  
Conversely, when participants encountered data irrelevant to the task, they preferred to use direct replacements. Names were frequently cited as data that could be safely replaced without affecting task outcomes. \blockquote[P3]{I'm gonna go for the names... I don't think that's relevant to this question.} These strategies underscore how participants actively engaged with the tool to tailor their privacy measures based on the specific task, data type, and expected response quality.

\subsection{Education and Learning Process}
During the testing, we observed that the extension played an educational role, shaping participants' mental models of data privacy and increasing their awareness of sensitive information.

\subsubsection{Educational Value of the Tool}
\label{sec:qual-findings-edu}
We summarize the educational values perceived by our participants below.

\textbf{Learning from the detection results (P2, P3, P4, P5, P6, P7, P11).}
The tool reminded users to reassess the data they share, as P7 mentioned,  \blockquote[]{the idea of highlighting it is a helpful thing. It lets you spot things that you may have put in there by accident, or the you may have just glossed over.}
Additionally, the tool helped participants recognize information they might not have considered sensitive. For example, P3 was surprised when ``Barnes and Nobles'' (the name of a book store) was flagged, but acknowledged it made sense when considering that there are not many of them in their location, which was specified in the prompt.
This highlights how the tool prompts users to identify unrecognized privacy risks.

    \textbf{Impact beyond one-time education (P2, P4, P5, P9, P11).}
    The tool's educational impact extended beyond one-time interactions. Participants’ understanding of sensitive data evolved with use. For instance, P5 experienced a shift in mental models after testing the extension across multiple sessions. They learned to mask certain terms like ``polyvagal therapist'' as they saw previous models flagged it and even when it was not flagged by a later model, suggesting the tool fosters awareness for safeguarding personal information. Even brief exposure to \systemname{} enhanced participants' ability to recognize sensitive data, indicating a lasting educational effect.

\subsubsection{Learning process of how to do sanitization}
Participants faced challenges with the abstraction feature, which was less intuitive at first. P2 admitted forgetting about abstraction, while P8 confused it with the replacement feature, expecting it to \blockquote{change it to more general terms.}
Additionally, people often lacked clear expectations of abstraction results, while they like to see things match their expectation.

Participants also experienced a learning curve in understanding the balance between safeguarding sensitive information and maintaining the quality of responses generated by \systemname{}. P2 doubted the system's ability to generate thorough answers after redacting health information. However, after testing, they found the sanitized input yielded similar-quality responses, realizing \blockquote{the substance [of the response before and after sanitization] isn't very different.} The evolution in participants' assessments suggests that while initial hesitation exists, confidence grows as users see that redaction doesn’t significantly impact response quality.

\section{Discussion}
We present discussions of our results, including the proposal of smaller-LLM-based privacy controls, validated by our studies as a promising solution to privacy issues in LLM-based systems; a summary of the value of user-led data minimization; and synthesized design implications for AI-powered, human-in-the-loop privacy controls.

\subsection{On the Feasibility of Smaller-LLM-Based Privacy Controls of LLM-Based Systems}
\label{sec:discussion-small-model-based-control}
Smaller LLMs (e.g., Llama3-8B) are widely considered to have the benefit of inherent privacy protection, as they can be deployed locally to avoid concerns associated with sharing data with larger LLMs' service providers.
Research and real-world applications have primarily focused on fine-tuning these smaller LLMs for specific downstream tasks (e.g., proofreading) to enhance their performance on those tasks.
In this work, we propose a different approach, leveraging smaller LLMs to empower privacy controls, particularly by helping users minimize unnecessary disclosure when interacting with LLM-based conversational agents.
Our prototyping and evaluation results have demonstrated the feasibility and potential of this idea in multiple aspects.

First, \modelone{} and \modeltwo{} achieved similar subjective perceptions regarding data minimization efficacy, reduced data sharing, privacy protection, and adoption intention (\autoref{sec:quant-findings-behavioral}).
Participants rated both versions of the tool favorably.
In our qualitative analysis (\autoref{sec:qual-findings-trust}), we attributed the varied preferences to different areas that the two versions of the system do well in.

We also observed a disparity between the subjective ratings, the objective reduction, and the accuracy benchmarking of the two models.
Although participants liked both versions, the actual reduction of disclosure when using \modeltwo{} is half of that of \modelone{} (\autoref{sec:quant-findings-behavioral}).
This observation corroborates the benchmarking of PII detection based on our taxonomy which reveals that GPT-4o outperforms Llama3-8B in recall (\autoref{sec:implementation-eval}).
This disparity is fundamentally related to users' limited experiences and established standards in the task. 
We observed that users did not evaluate the tool based on a comprehensive set of predefined ground truths for detection.
Instead, they were often prompted by our tool to become aware of sensitive disclosures that they had not realized before (\autoref{sec:qual-findings-edu}).
In this aspect, both models did a good job and significantly improved upon the status quo, as put by P5 when asked to compare the two versions, \blockquote{if I'm answering the question in a vacuum, and just comparing it against what ChatGPT originally provides, then it's a massive step up.}

The disparity between subjective ratings and objective reductions raises questions about how to prevent users from having a \textit{false sense of privacy}.
Specifically, how to address their over-reliance on the tool, with the assumption that it will completely remove their personal information, even though it may actually miss some.
One direction is to keep improving the accuracy of the small model for detection by fine-tuning the models or combining the results of multiple models.
Another direction is to calibrate users' expectations of the capability and scope of detection with the actual model's capability.
For example, we noticed that participants spontaneously compared the suggestions of the two models. When they encountered different results between the two models, they realized that one model did a better job than the other, and certain information was missing.
We may intentionally display results from multiple models to users to expose them to the potential risks of missed detection and reduce overreliance.

In addition to improvements to the small models, the response generation also needs enhancements to better accommodate the use of placeholders to generate accurate and correctly formatted responses, as indicated in our response utility analysis (see \autoref{sec:response-utility}).
Future work could explore introducing a stricter control mechanism during response generation.
For example, explicitly instructing the model to avoid creating new placeholders or to follow a predefined mapping schema could be beneficial. Additionally, using in-context learning to provide examples might prevent overly conservative answers when encountering placeholders.

Overall, these findings suggest that smaller LLMs, even without fine-tuning, already possess privacy-preserving capabilities to an extent that can be converted into enhanced privacy protection to end users.
However, room for improvement is also clear, particularly in terms of metrics that affect the user-led data minimization results.

\subsection{User-Led Data Minimization Bridges Gaps in Privacy Awareness and Controls}

Our study examines people's privacy concerns about using ChatGPT, focusing on the information identifiability and the lack of transparency.
Participants employed some data minimization strategies in their everyday usage, but found them inadequate, leaving them feeling vulnerable and lacking control (\autoref{sec:qual-findings-concerns}).
These findings validate the general need to offer users convenient and flexible support for data minimization and also generate deeper insights into how certain design choices of \systemname{} are useful to handle the nuanced and personal user risk perceptions.
For example, giving users full control over the sanitization actions was appreciated by users (\autoref{sec:qual-findings-trust-control}), and also helps them manage the privacy-utility tradeoffs following their preferences (e.g., varied assessment of identifiability in the same case, see \autoref{sec:qual-findings-concerns-identifiability} and \autoref{sec:qual-findings-tool-sanitization-strategies-datatype}).
Users valued our tool and were willing to adopt it especially because of many features we provided to streamline the process, such as providing alternative solutions (\autoref{sec:qual-findings-alternative-solutions}), convenience features like ``Select all'', and ``Write back'' (\autoref{sec:qual-findings-convenience-features}), and helping users find the middleground between privacy and utility by offering two options of sanitization targeting different scenarios, abstraction and replacement (\autoref{sec:qual-findings-tool-sanitization-strategies-actiontype}).

Another gap our tool addresses is increasing users' awareness of sensitive disclosure (\autoref{sec:qual-findings-edu}).
Participants frequently cited the tool's educational value as a benefit alongside enhanced privacy control and protection.
Although not specifically designed as an educational tool, our tool demonstrates an example of achieving privacy education by providing in-situ learning opportunities about the privacy risks in regular data-sharing, similar to the idea of experiential learning~\cite{gentry1990experiential}.
Users were willing to engage in exploring different sanitization options because it provided immediate and tangible benefits to them.
This insight could inform more privacy education design and research situated in the context of user-led data minimization.

\subsection{Implications for Designing AI-Powered Human-in-the-Loop Privacy Controls}

Our approach highlights the need for a shift from traditional human-out-of-the-loop Privacy-Enhancing Technologies (PETs) to tools that support human-in-the-loop, granular privacy management.
While AI is useful in achieving this goal, properly integrating AI into such systems, so that users will trust it, be willing to use it, and find that the protection it offers aligns with their expectations, requires an in-depth understanding of the dynamics of interactions between humans and AI, as moderated by the interfaces between them.
We reflect on lessons learned from our studies on designing such systems and interfaces below.

\textbf{Towards human-centered evaluation.}
Although performance metrics like accuracy, precision, and recall are all important ways to assess the performance of a model, their impact on user experience and decision-making varies significantly.
As highlighted in \autoref{sec:qual-findings-trust}, the comprehensiveness and consistency of the detection and sanitization results are key factors that affect users' trust.
Specifically, users were amenable towards false positives in detection results (e.g., less sensitive information) and mislabeled data categories, whereas a single false negative could be catastrophic to their trust in the system.
This suggests that certain metrics, like recall, are more important than others, like precision.

Additionally, we found that users' perceptions of sensitive disclosure span a wide range of topics, which are beyond the scope of our taxonomy and likely cannot be fully captured by any taxonomy (\autoref{sec:qual-findings-trust-complete-detection}).
When \modeltwo{} generated detection results outside the taxonomy (e.g. weight and height), users actually preferred them over \modelone{}, which more strictly adheres to the taxonomy and did not detect them as sensitive data.
However, common benchmarking methods driven by a taxonomy will misrepresent users' preference of this case in the result, which indicates the limitations of the taxonomy-based evaluation method and emphasizes the importance of privacy protection beyond PII~\cite{brown2022does}.

\textbf{Balance between comprehensiveness and control.}
In \autoref{sec:qual-findings-trust-control} and \autoref{sec:qual-findings-trust-complete-sanitization}, we observed seemingly conflicting needs concerning privacy controls.
We also saw in \autoref{sec:privacy-trade-off-balance} that users struggle to identify the optimal trade-offs between privacy and utility, due to a lack of a deep insights of the models' capabilities.
This tension highlights the need to strike a balance between giving users control to establish subjective trust and ensuring comprehensive data minimization without being undermined by users' limited experiences or attention spans.
Future research can further explore different levels of user engagement and agency, and the impact of individual difference, to design better interfaces that meet both important requirements.

\textbf{Consistency, predictability, and learning curve.}
Users reported abstraction being useful for retaining contexts while conveying general information.
However, it was used less often than expected, partly due to the lack of predictability in abstraction results.
This is exacerbated by the inconsistent results between partial and complete input (both versions) and across different trials (\modelone{} only).
As a result, the learning curve is heightened.
Additionally, predictability also involves aligning users' expected and actual abstraction granularity; overgeneralization is a main reason for dissatisfaction with abstraction results.

\textbf{Transparency and trust.}
Despite the positive feedback, users also expressed concerns about \blockquote[P4]{offsetting the privacy concern away from ChatGPT onto this extension}.
They stressed the importance of high transparency, expecting the tool to run locally.
This highlights that users may expect higher trust standards from privacy-protection tools.

\section{Conclusion}

In this work, we developed a system powered by LLMs that allows users to sanitize prompts, enabling user-led data minimization for LLM-based conversational agents. Our study revealed that most privacy concerns stem from the fear of being identifiable through their messages and the potential risks that follow. By offering quick, easy-to-use sanitization options, users could achieve satisfactory responses from GPT while feeling secure about their privacy. 
Notably, users became more aware of privacy considerations and actively evaluated whether exposing certain information was necessary for an appropriate response. Our findings also show no significant difference in user perception between the cloud-based and on-device models, suggesting that smaller LLMs offer a promising solution for on-device, user-facing privacy management in LLM-based applications.

\begin{acks}
The project is in part supported by a gift from Google on ``User-Centered Privacy Design in Smart Mobile Text Entry''.
 We are thankful to Weiyan Shi, Jackie Yang, Niloofar Mireshghallah, and all members of the PEACH lab for their helpful suggestions and feedback at different stages of this project.
\end{acks}

\bibliographystyle{ACM-Reference-Format}
\bibliography{sample-base}
\appendix

\definecolor{gray}{gray}{0.5}
\section{System Benchmarking and On-Device Feasibility Study}

\subsection{Time Performance Benchmark}
\label{sec:full-time-benchmark-result}

We present the full results of our experiments benchmarking the time performance of \modeloneintable{}, \modeltwointable{}, and Presidio.
For the on-device models, we ran the experiments with four varying device configurations, including three consumer computers: Apple M2 (24 GB memory), Apple M4 (16GB memory), Apple M1 Max (64GB memory), as well as one cloud virtual machine with an NVIDIA L4 GPU (24GB VRAM).

\begin{table}[h!]
\centering
\caption{Comparison of Response Times Among Presidio, \modeloneplain{} and \modeltwoplain{}. On-device models are benchmarked with different device configurations.}
\resizebox{\columnwidth}{!}{
\begin{tabular}{lcccccc}
\toprule
\textbf{Model} & \textbf{Device} & \textbf{Min (s)} & \textbf{Max (s)} & \textbf{Mean (s)} & \textbf{Std (s)} \\
\midrule
\modeloneplain{}  & & 0.53    & 3.98    & 1.41    & 0.65              \\
\modeltwoplain{} (first detection) & M2, 24GB &  1.34  &  4.52  &  2.01   &      0.48     \\
\modeltwoplain{} (full detection) & M2, 24GB & 1.62  &  7.93  &  3.40   &      1.29     \\
\modeltwoplain{} (first detection)  & M4, 16GB &  1.07 & 3.58 & 1.57 & 0.38       \\
\modeltwoplain{} (full detection)  & M4, 16GB &  1.23 & 6.35 & 2.66 & 1.02    \\
\modeltwoplain{} (first detection)  & M1 Max, 64GB &   0.45 & 1.60 & 0.66 & 0.16          \\
\modeltwoplain{} (full detection)  & M1 Max, 64GB &  0.54 & 2.58 & 1.09 & 0.41         \\
\modeltwoplain{} (first detection)  & L4, 24GB &   0.46 & 1.22 & 0.67 & 0.15          \\
\modeltwoplain{} (full detection)  & L4, 24GB &   0.52 & 2.87 & 1.16 & 0.47       \\
Presidio & M2, 24GB & 0.6 & 1.98 & 0.73 & 0.19     \\
Presidio  & M4, 16GB &   0.48 & 0.82 & 0.53 & 0.06  \\
Presidio  & M1 Max, 64GB &  0.74 & 9.64 & 0.84 & 0.53  \\
Presidio  & L4, 24GB &   1.56 & 8.43 & 1.73 & 0.47       \\
\bottomrule
\end{tabular}
}
\label{tab:full_response_time_table}
\end{table}

\subsection{On-Device Feasibility Study Information}
We collected user demographic and device information to ensure participants met the study requirements and represented a diverse range of backgrounds and device configurations. See \autoref{tab:on-device-sample-info} for more detail.
\label{Ondevice}
\begin{table}[htbp]
\centering
\caption{Participant and Device information of the On-Device Feasibility Testing Study}
\resizebox{\columnwidth}{!}{%
\begin{tabular}{c c c c c c c}
\toprule
\textbf{ID} & \textbf{Gender} & \textbf{Age Group} & \textbf{Background} & \textbf{Chip} & \textbf{Memory (GB)}  \\ 
\midrule
F1  & Male & 25-34 & Tech & M1 Max  & 64  \\ 
F2  & Male & 25-34 & Tech & M1 Pro  & 16 \\ 
F3  & Female & 18-24 & Non-tech & M1 Pro  & 16 \\ 
F4  & Female & 18-24 & Non-tech & M2  & 16 \\ 
F5  & Female & 18-24 & Non-tech & M4  & 16 \\ 
F6  & Female & 18-24 & Non-tech & M1  & 16 \\ 
F7  & Female & 25-34 & Mixed & M1  & 16 \\ 
F8  & Female & 18-24 & Mixed & M3 Pro  & 36 \\ 
\bottomrule
\end{tabular}
}
\label{tab:on-device-sample-info}
\end{table}

\section{Interview Protocol}

\subsection{Pre-Interview Message}

Thank you for your willingness to join our study! During the study, we will test our privacy-preserving extension together. Specifically, the researcher will ask you to provide real-time feedback while you use our extension during a conversation with ChatGPT.

\textcolor{gray}{(Make sure the participant received the email with instructions and survey link.)}

Here are a few things to set you ready!

\begin{enumerate}
    \item Please fill out the survey so we can better prepare for your experience. It should take less than 3 minutes to finish. 
    \item Please prepare a ChatGPT prompt to test beforehand. We highly suggest you review your ChatGPT chat history and pick a past conversation that:
    \begin{itemize}
        \item Contains personal information that you are concerned about sharing with ChatGPT.
        \item You are comfortable showcasing for the study purpose.
    \end{itemize}
    \item Feel free to take a look at the participant information sheet for more information about the study. All your personal information will be kept confidential.
\end{enumerate}

\textcolor{gray}{(If the participant has questions about how to select a ChatGPT conversation, provide guidance based on their concerns about privacy.)}

\subsection{Counterbalancing Settings}
The study involves 12 participants across 4 conditions, where each participant interacts with two models (Model 1 and Model 2), using both their own and provided examples. The order of testing is randomized to avoid bias. See Table ~\ref{table1}.

\begin{table}[htbp]
\centering
\caption{Counterbalancing settings}
\resizebox{\columnwidth}{!}{%
\begin{tabular}{|c|c|c|c|}
\hline
\textbf{ID} & \textbf{Example} & \textbf{First Example} & \textbf{Model Order} \\ 
\hline
P1  & E1  & Participant’s Example & Model 1 $\rightarrow$ Model 2 \\ 
P2  & E1  & Participant’s Example & Model 2 $\rightarrow$ Model 1 \\ 
P3  & E2  & Participant’s Example & Model 1 $\rightarrow$ Model 2 \\ 
P4  & E2  & Participant’s Example & Model 2 $\rightarrow$ Model 1 \\ 
P5  & E3  & Participant’s Example & Model 1 $\rightarrow$ Model 2 \\ 
P6  & E3  & Participant’s Example & Model 2 $\rightarrow$ Model 1 \\ 
P7  & E1  & Experimenter’s Example & Model 1 $\rightarrow$ Model 2 \\ 
P8  & E1  & Experimenter’s Example & Model 2 $\rightarrow$ Model 1 \\ 
P9  & E2  & Experimenter’s Example & Model 1 $\rightarrow$ Model 2 \\ 
P10 & E2  & Experimenter’s Example & Model 2 $\rightarrow$ Model 1 \\ 
P11 & E3  & Experimenter’s Example & Model 1 $\rightarrow$ Model 2 \\ 
P12 & E3  & Experimenter’s Example & Model 2 $\rightarrow$ Model 1 \\ 
\hline
\end{tabular}
}
\label{table1}
\end{table}

\subsection{Interview Script}

Hi there! We are researchers from Northeastern University, and we are interested in learning about your privacy concerns when using ChatGPT. During this interview, we will test a privacy-preserving extension together. Specifically, I will ask you to provide feedback while you use our extension during conversations with ChatGPT.

\textcolor{gray}{(Send the information sheet to the participant before starting the conversation.)}

Here is the study information sheet. During the study, you are free to answer only the questions you're comfortable with. If there’s a question that you don’t want to discuss, just let me know. It won’t affect your compensation.

\textcolor{gray}{(Ensure the participant understands the compensation method: Amazon gift card or Prolific payment.)}

I will also be video recording this meeting for note-taking purposes. All your responses will be kept confidential. You are welcome to turn off your camera anytime during the study. Does that work for you?

\textcolor{gray}{(If they answer yes, start recording the session.)}

Great! Let’s start.

\subsection{General Questions}

First, I would love to know what you usually use ChatGPT for. Could you give a recent typical example?

\textcolor{gray}{(Note the participant's response and be ready to probe further based on their usage habits.)}

\textcolor{gray}{[If they have tried to hide PII before]} I saw in the survey that you selected “yes” for trying to minimize unnecessary information shared with ChatGPT, would you mind explaining more about how you took what action under what cases? 

\textcolor{gray}{[If they have not tried to hide PII before]} I saw in the survey that you are concerned with personal information shared with ChatGPT, could you explain more about the cases?

That’s a great example! Perfect, our team has created a tool, which is a Chrome extension that would be useful for identifying your private information and masking it with general terms while using GPT. We are honored to invite you to try it out and give feedback through my following instructions.

We hope to have you test the extension with a real-world conversation that you had with ChatGPT that involves sharing of your personally identifiable information, your personal experiences, or other sensitive information. If you have prepared the conversation before the study, that’s great and you can directly use that in the testing. If not, no worries. You can take a few minutes to review your ChatGPT conversation history and pick one that you feel comfortable sharing with us.

\textcolor{gray}{Let user copy and paste the example to the blank doc page so that we can all review it.}

Great! Could you show the conversation to us and briefly explain why you picked it?

Thanks! Now we’ll play a demo video to show you our extension works:

\textcolor{gray}{Play demo video}

\subsection{Contextual Experience User Testing}

We designed two versions of the systems to test the efficacy of different models for the privacy information masking task. To avoid biasing you when testing them, we will share details of how the two models work at the end of the interview.

We would love you to test 2 conversations with 2 different models of the extension, so in total, 4 sessions with ChatGPT using our extension. One is the real-world conversation you prepared and just demonstrated to us; another is an example we prepared. Please interact with ChatGPT as you would do in a realistic setting, including asking follow-up questions to fulfill your request. We are all ears to hear about your feelings, including both positive and negative feedback. These will help us to iterate.

Session 1: Let’s start with your personal example/our hypothetical example. You can find the conversation you prepared and use it to test the extension. Feel free to ask questions as we go.

\textcolor{gray}{(Guide participant through the process of using the extension for their first conversation.)}

Session 2: Now, let’s switch to the other model and try the same example again.

\textcolor{gray}{(Ensure the participant understands they need to test the same conversation with the other model.)}

Session 3: Now, let’s switch to our hypothetical example/your personal example. I will send the prompt over to you. Feel free to ask questions if anything is unclear.

\textcolor{gray}{Choose one example from E1/E2/E3 based on the counterbalacing settings table}

Session 4: Finally, let’s test the hypothetical example again using the other model.

\subsection{Feedback Questions For Each Session}

Based on your experience using the extension, I have a few questions:
\begin{itemize}
     \item How satisfied are you with the response after using the extension? 
    \item To what extent do you believe this experience helps protect your privacy? Why?
    \item What do you like or dislike about this experience? Why?
\end{itemize}

\textcolor{gray}{(Note any interesting behaviors or hesitations and ask follow-up questions as necessary.)}

\subsection{Post-Session Survey}

Please fill out this survey based on your experience: [survey link removed]

Please choose [A/B] for setting 1 and [E1/E2/E3/E4] for setting 2.
\begin{itemize}
    \item A: Cloud-based (A)
    \item B: On-device (B)
    \item E1: Experimenter's example 1
    \item E2: Experimenter's example 2
    \item E3: Experimenter's example 3
    \item E4: Participant's own example
\end{itemize}

\subsection{Model Comparison After Session 2 And Session 4}

Thanks for filling out the surveys! Let’s briefly go over your responses for both models:
\begin{itemize}
    \item The extension reduces the disclosure of unnecessary information to ChatGPT.
    \item The extension reduces the disclosure of my personal information to ChatGPT.
    \item I have fewer privacy concerns when using this extension.
    \item I would use this extension if it were well-designed for usability.
\end{itemize}

\textcolor{gray}{[If they gave the same score for both models]} I noticed you gave the same score for both models. Could you explain why?

\textcolor{gray}{[If they gave different scores]} I noticed you gave different scores for the models. Could you explain the differences you observed?

\subsection{Debrief}

We designed the two versions of the systems to test the efficacy of different models for the privacy information masking task. We didn't share the details of how the two models work to avoid biasing you when answering these questions. Since you have finished all the tasks, I would like to share more information about the two models we used today.

In Model 1, personal information detection is implemented by prompting GPT-4. In Model 2, personal information detection is implemented by prompting a smaller large language model hosted on our server. In the GPT-4 powered model, OpenAI may securely retain your data for up to 30 days, and it will not use your data for training their models. In the self-hosted model, we don’t store any data that you share during the study. If there is certain information you would like us to delete, please feel free to contact us and we will make our best efforts to delete it.

Do you have any questions?

\subsection{Closing Remarks}

That's a wrap for our interview today! Thank you so much for taking the time to share your insights and experiences with us - it's been incredibly valuable. 
We hope you find it useful. Please don't hesitate to reach out if you have any further questions or thoughts you'd like to share. Thanks again for your contribution to our study!

\section{Qualitative Code Book}
\label{sec:codebook}

\captionof{table}{Qualitative Code Names and Sample Quotes}
\begin{supertabular}{|>{\raggedright}p{0.2\linewidth}|p{0.3\linewidth}|p{0.4\linewidth}|}
\hline
\textbf{Code Name} & \textbf{Definition} & \textbf{Quotes} \\
\hline

Feature: Replace Back & Replace placeholder text with PII in both messages sent by users and responses generated by GPT. & \textit{Because realistically, when you're putting in personal information about yourself and you need it contextualized for specific use, even if it's something as simple as a name for readability and things like that}\\ \hline
Feature: Revert Option & Users appreciate the revert button, which allows them to return the text to its original state. & \textit{Afterwards, mask [any identifying information] for ChatGPT, but then re-transcribe [ChatGPT's response] back using your own identifiers. So you're able to follow the conversation, and ChatGPT does not have access to the identifying information that you would give it.}\\ \hline
Feature: PII Highlighting & Highlight all PIIs detected & \textit{I think it's nice that, you know, it's presenting it to you in a way that highlights what it considers to be a problem.}\\ \hline
Feature: Select All Option & Select all PIIs detected and then redact & \textit{I don't need that [one to be] replaced, but I'm going to just do a select all [for convenience].}\\ \hline
Actionable alternatives & Our tool offers alternative solutions through actionable suggestions. & \textit{It identifies it, and it also provides solutions instead of just marking a word.} \\ \hline
User Control And Flexibility  & Flexibility and control offered by the tool. & \textit{I'm satisfied with the control that I have of the extension.} \\ \hline
Unexpected Detection & The tool increases awareness by detecting things people were not expecting. & \textit{It's also a pretty cool, educational experience for myself.} \\ \hline
Unexpected Sanitization Suggestion & The tool increases awareness by suggesting sanitization (replacement or abstraction) users were not aware of. & \textit{It's a sense of educational experience. I wouldn't have thought that I could just like input and stuff like ADDRESS1, NAME1 that's really useful instead of me having to take that out each time it it does it.} \\ \hline

Process Convenience & The tool offers features that make the whole process effective. & \textit{Having that ability to just quickly sanitize it and then have it feed out what you want it to, and then have the extension replace those terms back from what ChatGPT works with, seems like it would be very helpful in terms of using names and stuff like that, and like location data without like having to actually feed your information in.} \\ \hline
Convenience - Grouping of Items of Same Label & The tool groups items that are detected with the same label in suggestions. & \textit{Removing everybody's name. I do like that. It aggregates all the names right away.} \\ \hline
Complete Redaction & The tool makes it easy for users to completely redact all the items they intend to deal with. & \textit{I really enjoyed that everything was sorted properly, and that it was a lot easier for me to go through check boxes and know that I'm not missing one because they're all there.} \\ \hline
Redaction Incompleteness & The tool lacks in providing ways to help users redact all the intended items. & \textit{I do notice that there's a floating Martin right here. I don't know how that stuck around.} \\ \hline
Missing Fail-Safe Mode & The tool lacks a fail-safe mode (forcing users to check before sending a response), and users want higher security. & \textit{If I do need to use the tool to do something truly sensitive, it would be nice to know that there's a safeguard in place that I'm not accidentally pressing Enter key.} \\ \hline
Workflow Disruption & Using the tool requires users to change their regular workflow with ChatGPT. & \textit{On a product standpoint: It's a shame that I would have to change the way that I operate with ChatGPT, because my instinct is to treat it like a chatbot, just like typing input in order to get the stuff that I want out of it.} \\ \hline
Missing Abstraction Visual Differences & The tool lacks visual indication to show which parts have been changed after abstraction. & \textit{So I'm not sure, other than the highlighting, how much of a difference I see in what it's doing?} \\ \hline
Abstraction Doesn't Support Replace-Back & Unlike replacement, abstraction doesn't support replacing back. & \textit{Although what is interesting is when I submitted this, it didn't revert back in the response. Like, right here, in the highlighted specific therapy type—that was something that I abstracted. But in this version, it doesn't show up in green for what my specific therapy type was, and I think that would be a little bit confusing to me.} \\ \hline
Privacy Concern: Risk and Harm & People worry about potential risks of using GPT. & \textit{If there was some sort of data leak like if that could be used to help identify me or know where I do my banking. Let's say, a potential scammer, or some person try to bruteforce their way into an account that I might have at this particular bank, that could be problematic.} \\ \hline
Privacy Concern: Uncertainty & People worry about using GPT because its lack of transparency. & \textit{For me, the privacy concerns are heightened because we just don't know what the capabilities of this are, so I don't know what I'm putting information into. It reminds me of like when you're, you know, talking to even like your credit card company on the phone, and they ask for your social security number. Like, I don't know where that's going, and that is the same way that my privacy concerns are heightened here.} \\ \hline
Privacy Concern: Identifiability & People worry about using GPT because they are concerned they will be identifiable. & \textit{Maybe this is just a personal preference, but I do think that 6 foot 3 and weigh 200 pounds are identifiers that could be more general. But when they're all taken together, I think that it paints a different picture.} \\ \hline
Privacy Concern: Related People's Identifiability & People worry about using GPT because they are concerned people related to them will be identifiable as well. & \textit{I have an employee who had a baby with his unmarried girlfriend. I chose to say friend instead of employee, because again, it seems sensitive.} \\ \hline

Privacy Concern: Legal Issues & People worry about using GPT because of legal consequences once they are identifiable. & \textit{OpenAI, you have this data on this person, and we're going to subpoena. They're gonna subpoena all this information of me, like usage history with ChatGPT.} \\ \hline
Privacy Concern: Personal vs. Professional & People have concerns about GPT related to the personal vs. professional settings. & \textit{I'm a novelist and an editor, you know, I'll tell it that, but, like, I don't really want it knowing, like, my health stuff or stuff about my personal life.} \\ \hline

Privacy Concern: Learning Ability & People worry that GPT will use their info to train itself. & \textit{I think the biggest concern that I had is that, you know, if we're inputting information into this, and it's learning about it. I wouldn't want it to learn specifically about me, if that makes sense. And so I've changed names. I've changed dates. Things like that.} \\ \hline

Privacy Concern: Distrust in Big Tech & People don't trust big tech behind GPT. & \textit{All of these large language models, like, most of them are run by big tech, and I don't trust big tech.} \\ \hline

Privacy Concern: Lack of Control & People worry that once they send their info to GPT, there is no way to control/safeguard it anymore. & \textit{I assume that I could go through and delete it. But like I'm already under the mindset that like when you delete something on somebody else's database, it's not gone.} \\ \hline

Privacy Concern About Data Type: Confidential & People worry about sharing confidential information like passwords with GPT.  & \textit{I guess there would be a hierarchy of things that would be like sensitive data that shouldn't be written out like, obviously, if people share like passwords or stuff with ChatGPT. That would be sensitive data, things that shouldn't be stored in external server that I have no control over.} \\ \hline

Privacy Concern About Data Type: Identifiable Data & People worry about sharing data like name that can be used for identification with GPT. & \textit{I think in terms of what I would find to be most useful in a conversation like this, specifically, it would be that, you know, because a lot of the personal information here is just your name and you can't really abstract the name.} \\ \hline

Privacy Concern About Data Type: Nonidentifiable But Sensitive & People feel uncomfortable sharing non-identifiable but sensitive information such as beliefs, religion, and health data. & \textit{I had a conversation with it that I wasn't necessarily super comfortable with. I did that recently a couple of conversations having to do with beliefs.} \\ \hline

Privacy Concern About Data Type: Possessive Terms & People do not want to share sentences that include ``my'', ``I''. & \textit{Because environmental engineering is generic. But would someone want to say, hey, ``my background'', [it] might trigger your algorithm.} \\ \hline

Privacy Paradox & People are concerned about the privacy issue with using GPT, but still use it. & \textit{I had a conversation with it that I wasn't necessarily super comfortable with. I did that recently, a couple of conversations having to do with beliefs.} \\ \hline

Data Minimization Strategy - Reacting To The Tool  & How people finally decide what to minimize based on the tool's suggestion, redaction and abstraction results. & \textit{(Reviewing the list of suggestions) So I think the things that I would want to obscure here that I don't think would change the efficacy of the response would be time, the health information, the type, and everything else seems fairly general to me.} \\ \hline

Data Minimization Strategy - Tradeoff & People made trade-offs as they perceived the response would not be as good without exposing some personal details. & \textit{I feel like, if I choose to replace it, [ChatGPT] will obviously not be able to say anything.} \\ \hline

Learning Process of Data Minimization - Learning Impact of Minimization & People learn the extent they can minimize by seeing the response GPT generates. & \textit{I think I'm satisfied, you know, like I say, it doesn't, it doesn't seem to be hugely like, it doesn't seem to be as detrimental as I thought.} \\ \hline

Learning Process of Data Minimization - Trial N Error & People learn whether they prefer abstraction or replacement through trial and error. & \textit{I think what I would want to do is go back and actually abstract that one.} \\ \hline

Differing Minimization Standards & People decide to not follow what is suggested by the tool. & \textit{I don't think that it minimizes the necessary information. I think specifics like duration of therapy are more crucial than the gender identification of the person that I was talking to.} \\ \hline

Preference for Possessive Terms Minimization & People wants to replace possessive terms like ``I'', ``my'' even though it is not suggested by the tool. & \textit{Environmental engineering is generic. But would someone want to say ``my background''? ``my background'' might trigger your algorithm.} \\ \hline

Domain-Specific Minimization & People's minimization method depends on their own domain knowledge and the domain knowledge of the message. & \textit{Obviously, when it comes to law, some of it is location based, so you might have some interesting things that pop up when it tries to generalize county and stuff like that.} \\ \hline

Conservative Minimization Approach & People could be conservative when minimizing unnecessary information. & \textit{I think I'm satisfied, you know, like I say, it doesn't seem to be as detrimental as I thought.} \\ \hline

Existing Minimization Habits & People already have the habit of minimizing info themselves, so they do not want to adopt the tool. & \textit{It might be kind of cool to, like, highlight stuff that maybe I didn't think of, you know, but the steps to to actually use the extension might be, I don't know, a little bit more inconvenient that then more trouble than what it's worth when I can just probably write around it anyway.} \\ \hline

Redaction Completeness Affects Trust & Incomplete redaction affects user trust in the tool. & \textit{You know, again, when you're looking from the perspectives of privacy, you would want to remove every instance of whatever it is that you think is information you don't feel like sharing with ChatGPT.} \\ \hline

Abstraction Consistency Affects Trust & Inconsistent abstraction results affect user trust in the tool. & \textit{It's gonna change to ``my city'' now, which is different from ``major city in the Midwest''. But also, in my mind, this opens up something else, which is that if it says in ``my city'', that dramatically changes the integrity of the answer that I'm gonna get.} \\ \hline

Good Abstraction: Context Maintained & Good abstraction generalizes while maintaining context.  & \textit{I would select that, and abstract, and let's see what it changed it to. The extension changes ``KCMO'' to ``a major city  in the Midwest''.} \\ \hline

Good Abstraction Increases Trust & Good abstraction leads to user trust in the tool. & \textit{I think, again, the piece for me, for the second model that seems very encouraging, is that it takes into account local resources and the climate.} \\ \hline

Unsatisfactory Abstraction: Lacks Specificity & Unsatisfactory abstraction keeps some context but is not specific enough. & \textit{So in my original response, before I was inputting stuff into this study, it came up with information about like the climate and the place and things like that. And I'm just curious about location ``in a major city in the Midwest''. Like I'm curious if that is even too general.} \\ \hline

Over-Abstracted: Lost Context & Bad abstraction removes too much detail, resulting in loss of context. & \textit{But let's see what it does with the abstract. OK, jurisdiction. I can imagine the issue that's gonna arise there, but I will see what it outputs. OK, so it just goes far more general.} \\ \hline

Nature of Abstraction: Hard to Generalize & The nature of abstraction means it sometimes removes too much detail, even when humans might struggle to abstract well. & \textit{I feel like ``partner'' might abstract it too much, and wouldn't necessarily get me what I need.} \\ \hline

Bad Abstraction: Increased Sensitivity & Bad abstraction replaces PII with something even more sensitive. & \textit{That [``unplanned pregnancy''] almost seems worse than than what it [``have a baby''] is . Actually, I feel that ``unplanned pregnancy'' has its own connotations and its own privacy concerns as well, like it's related to the topic of the person itself.} \\ \hline

Abstraction: Need Time to Learn & People are not that familiar with how abstraction works and need time to learn. & \textit{Because I forgot about it, if I'm being honest.} \\ \hline

Abstraction: Different From Expectation & The abstraction result is different from what the user expectated. & \textit{And I said abstract Skype. Did it just take it out? OK. I mean that makes sense.Not saying it's necessarily the wrong approach, but I guess it's not necessarily the thing you expect when he used that button. } \\ \hline

Inconsistent Detection:Trust Issue & Inconsistent detection suggestions lower user trust in the tool. & \textit{Skeptical knowing that it flagged this the second time, but not the 1st time. Again, it loses a little bit of credibility with me that this experience is making me skeptical of it, and thinking, oh, this is not that robust of a tool.} \\ \hline

False Positives: Tolerance & Detecting more information that’s considered less sensitive is acceptable. & \textit{I don't see it [``therapist''] as a name. That's just a general term, but I wouldn't be worried about that.} \\ \hline

False Negatives: Trust Issue & Detecting less information than expected is not acceptable. & \textit{What about the stuff it doesn't catch? Like, hey, it didn't catch everything. So why am I using this? I'm going to go back and revise it, you know? So that's like twice the work.} \\ \hline

False Negatives: Convenience Issue & Detecting less information than expected brings more inconvenience for users. & \textit{What about the stuff it doesn't catch? Like, hey, it didn't catch everything. So what am I using this? And then I'm going to go back and revise my thing as it is, you know? So that's like twice the work.} \\ \hline

Detecting More Sensitive Data: Trust Increase & Detecting more info than expected is good. & \textit{It's nice to know that—I mean, I guess, obviously, I would have known that ``unmarried girlfriend'' is an identifier, is a privacy thing} \\ \hline

Better Detection Without Following Taxonomy & Sometimes, the tool doesn’t follow the instruction well, but it leads to better detection beyond the taxonomy. & \textit{It identified things that I didn't think were personal identifiers. It highlighted my height, my weight, as well as the location that I was in, and I think that was really important to see that didn't show up in the 1st exercise.} \\ \hline

Data Categorization, Labeling Error & Mis-categorization could lead to lower user satisfaction. & \textit{I think that might be an incorrect analysis of what it is, but I see the difference.} \\ \hline

Low Expectations for Non-Identifiable Detection & People did not expect that non-identifiable information could be detected before using the tool. & \textit{Because, like, this whole conversation for me was dicey. For you guys, it's not, like, personally identifiable, but like, for me, this was kind of like, I don't know it was weird for me discussing this with a bit, because I kind of don't want it to know that I do meditation, or why.} \\ \hline
Transparency of The Tool & Transparency of the tool leads to more user trust. & \textit{That's just a matter of transparency on an extension like this, where it, I mean again, most of the time extensions aren't gonna be collecting the information, but you never know that they're not.} \\ \hline

Trust of the Tool versus ChatGPT & People debate whether to trust the tool or GPT. & \textit{You're trusting a different thing to protect your privacy. You're essentially trusting an extension to protect your privacy as opposed to ChatGPT.} \\ \hline

On-Device Processing & The tool's on-device capability increases transparency and user trust. & \textit{I think this is excellent for protecting the privacy. I assume that the extension is running itself locally and is not storing any of the data itself.} \\ \hline

Fast Detection Speed & The tool offers fast detection speed. & \textit{It's honestly a lot faster than I thought it would be in terms of how quickly it outputs its response.} \\ \hline

\end{supertabular}

\section{Hypothetical Examples}
\label{sec:hypothetical-examples}
\subsection{E1}
Please read the following email and write an appropriate response, I am willing to cooperate:

\noindent {Emily Clarke  \textless emily.clarke@templestate.edu\textgreater \\
To:\\
martinj@silverplan.gov.us\\
Wed 06/26/2024 13:05\\ \\
Hi Martin,\\}
I hope this email finds you well. I recently attended your presentation on Sustainable Urban Development and was thoroughly impressed by your insights and the innovative approach you're taking. I believe that there is a great potential for collaboration between us, especially considering my background in Environmental Engineering.

I'd love to discuss this further and see how we can work together on your project. Additionally, I have a few suggestions that might enhance your current approach:
\begin{enumerate}
    \item Consider integrating green roofing systems to improve energy efficiency.
    \item Incorporate more community engagement in the planning stages.
    \item Utilize data analytics to monitor and optimize resource usage.
\end{enumerate}
Looking forward to your thoughts on this. Let's schedule a meeting soon to discuss this in more detail.
{\\ \\Best Regards, \\ Dr.Emily Clarke \\ Assistant Professor \\ Temple State University \\ TRB 113}

\subsection{E2}
Review the following dataset and come up with insightful observations:

Trip ID	Destination	Start date	End date	Duration (days)	Traveler name	Traveler age	Traveler gender	Traveler nationality	Accommodation type	Accommodation cost	Transportation type	Transportation cost

1	London, UK	5/1/2023	5/8/2023	7	John Smith	35	Male	American	Hotel	1200	Flight	600

2	Phuket, Thailand	6/15/2023	6/20/2023	5	Jane Doe	28	Female	Canadian	Resort	800	Flight	500

3	Bali, Indonesia	7/1/2023	7/8/2023	7	David Lee	45	Male	Korean	Villa	1000	Flight	700

4	New York, USA	8/15/2023	8/29/2023	14	Sarah Johnson	29	Female	British	Hotel	2000	Flight	1000

5	Tokyo, Japan	9/10/2023	9/17/2023	7	Kim Nguyen	26	Female	Vietnamese	Airbnb	700	Train	200

6	Paris, France	10/5/2023	10/10/2023	5	Michael Brown	42	Male	American	Hotel	1500	Flight	800

7	Sydney, Australia	11/20/2023	11/30/2023	10	Emily Davis	33	Female	Australian	Hostel	500	Flight	1200

8	Rio de Janeiro, Brazil	1/5/2024	1/12/2024	7	Lucas Santos	25	Male	Brazilian	Airbnb	900	Flight	600

9	Amsterdam, Netherlands	2/14/2024	2/21/2024	7	Laura Janssen	31	Female	Dutch	Hotel	1200	Train	200

10	Dubai, United Arab Emirates	3/10/2024	3/17/2024	7	Mohammed Ali	39	Male	Emirati	Resort	2500	Flight	800

11	Cancun, Mexico	4/1/2024	4/8/2024	7	Ana Hernandez	27	Female	Mexican	Hotel	1000	Flight	500

\subsection{E3}
hey there! can you help me write a thank you letter to my therapist from the treatment center i was at for 35 days? she is a polyvagal therapist, and she has taught me so much. i want her to know how grateful i am. 

for backstory, i’ve been in therapy for half of my life. i intellectualize everything, and she has met me with understanding, grace, compassion, tough love when i needed it. i run away from places, painful feelings, and i am a person who lives in my head, and not my body. we discovered together that i feel like my legs are unsafe, and has really opened a door to me. i’ve had many therapists in my life, but she is incredible and has made a lifelong impact on me. i will forever be grateful

\section{Prompts}

\subsection{Detection}
\noindent You are an expert in cybersecurity and data privacy. You are now tasked to detect PII from the given text, using the following taxonomy only:

\begin{description}
    \item [\textbf{ADDRESS}]
    \item [\textbf{IP\_ADDRESS}]
    \item [\textbf{URL}]
    \item [\textbf{SSN}]
    \item [\textbf{PHONE\_NUMBER}]
    \item [\textbf{EMAIL}]
    \item [\textbf{DRIVERS\_LICENSE}]
    \item [\textbf{PASSPORT\_NUMBER}]
    \item [\textbf{TAXPAYER\_IDENTIFICATION\_NUMBER}]
    \item [\textbf{ID\_NUMBER}]
    \item [\textbf{NAME}]
    \item [\textbf{USERNAME}]
    \item[\textbf{KEYS:}] Passwords, passkeys, API keys, encryption keys, and any other form of security keys.
    \item[\textbf{GEOLOCATION:}] Places and locations, such as cities, provinces, countries, international regions, or named infrastructures (e.g., bus stops, bridges, etc.).
    \item[\textbf{AFFILIATION:}] Names of organizations, such as public and private companies, schools, universities, public institutions, prisons, healthcare institutions, non-governmental organizations, churches, etc.
    \item[\textbf{DEMOGRAPHIC\_ATTRIBUTE:}] Demographic attributes of a person, such as native language, descent, heritage, ethnicity, nationality, religious or political group, birthmarks, ages, sexual orientation, gender, and sex.
    \item[\textbf{TIME:}] Description of a specific date, time, or duration.
    \item[\textbf{HEALTH\_INFORMATION:}] Details concerning an individual's health status, medical conditions, treatment records, and health insurance information.
    \item[\textbf{FINANCIAL\_INFORMATION:}] Financial details such as bank account numbers, credit card numbers, investment records, salary information, and other financial statuses or activities.
    \item[\textbf{EDUCATIONAL\_RECORD:}] Educational background details, including academic records, transcripts, degrees, and certifications.
\end{description}

For the given message that a user sends to a chatbot, identify all the personally identifiable information using the above taxonomy only, and the entity\_type should be selected from the all-caps categories. 

Note that the information should be related to a real person not in a public context, but okay if not uniquely identifiable.

Result should be in its minimum possible unit.

Return me ONLY a JSON in the following format:
{\texttt{\raggedright \{"results": [\{"entity\_type": YOU\_DECIDE\_THE\_PII\_TYPE, "text": PART\_OF\_MESSAGE\_YOU\_IDENTIFIED\_AS\_PII\}]\}}}

\subsection{Abstraction}
Rewrite the text to abstract the protected information, without changing other parts. 

For example:

Input: <Text>I graduated from CMU, and I earn a six-figure salary. Today in the office...</Text><ProtectedInformation>CMU, Today</ProtectedInformation>
        
Output JSON:
\texttt{\raggedright \{"results": [\{"protected": "CMU", "abstracted": "a prestigious university"\}, \{"protected": "Today", "abstracted": "Recently"\}]\}}
Please use "results" as the main key in the JSON object.
\subsection{Response Comparison}
\label{sec:response-comparison-prompt}
Please act as an impartial judge and evaluate the quality of the responses to the user's message below. Choose the assistant that better follows the user's instructions and addresses the message effectively. Consider factors such as helpfulness, relevance, accuracy, depth, creativity, and detail. Ensure your evaluation is unbiased and not influenced by the order, names, or length of the responses.

Divide your evaluation into two parts: \textbf{format} (clarity, structure, and readability) and \textbf{content} (quality, accuracy, and relevance). Judge each part separately to avoid overlap. Only assign a far better score (\texttt{[[1]]} or \texttt{[[5]]}) if one response is significantly better in meeting the user's needs, showing notable improvements or insights that the other clearly lacks. Slight advantages should be scored as \texttt{[[2]]} or \texttt{[[4]]}, and similar responses as \texttt{[[3]]}.

Provide your final verdict in the following format:
\linebreak
\parbox{\columnwidth}{\texttt{\raggedright \{ "format\_score": "[[1]]" or "[[2]]" or "[[3]]" or "[[4]]" or "[[5]]", "content\_score": "[[1]]" or "[[2]]" or "[[3]]" or "[[4]]" or "[[5]]", "explanation": "Your explanation here in no more than 50 words." \}}}

Use these scoring guidelines:
\begin{itemize}
    \item [\texttt{[[1]]}]: Response A is far better.
    \item [\texttt{[[2]]}]: Response A is slightly better.
    \item [\texttt{[[3]]}]: Both responses are similar.
    \item [\texttt{[[4]]}]: Response B is slightly better.
    \item [\texttt{[[5]]}]: Response B is far better.
\end{itemize}

Far better (\texttt{[[1]]} or \texttt{[[5]]}) means a response provides a substantial improvement, e.g., critical errors corrected, significantly higher detail, or much better alignment with the user's instructions. For minor differences, choose \texttt{[[2]]} or \texttt{[[4]]}. Do not include any additional text, comments, or formatting outside the JSON structure.

\section{Response Comparison Examples}
\label{sec:response-comparison}

\begin{table}[htbp]
\scriptsize
\setlength{\extrarowheight}{1pt}
\setlength{\tabcolsep}{2pt}
\centering
\caption{E1 Responses Comparison with \modeloneintable{} (Format Score: original is slightly better, Content Score: original is far better)}
\renewcommand{\arraystretch}{0.95}
\begin{tabularx}{\columnwidth}{|X|X|X|}
\hline
\textbf{Redacted Input} & \textbf{Original Output} & \textbf{Replaced-Back Output of Redacted Message} \\
\hline
Please read the following email and write an appropriate response: \newline
\newline
[Sender's Name and Contact Information Redacted] \newline
To: [Recipient's Contact Information Redacted] \newline
[Date and Time Redacted] \newline
Hi [Recipient's Name Redacted], \newline
I hope this email finds you well. I recently attended your presentation on Sustainable Urban Development and was thoroughly impressed by your insights and the innovative approach you're taking. I believe there is a great potential for collaboration between us, especially considering my background in Environmental Engineering. I'd love to discuss this further and see how we can work together on your project. \newline
Additionally, I have a few suggestions: \newline
- Consider integrating green roofing systems to improve energy efficiency. \newline
- Incorporate more community engagement in the planning stages. \newline
- Utilize data analytics to monitor and optimize resource usage. \newline
Looking forward to your thoughts on this. Let's schedule a meeting soon to discuss this in more detail. \newline
Best regards, \newline
[Sender's Title and Affiliation Redacted]
&
{Subject: Re: Collaboration on Sustainable Urban Development \newline
Hi Dr. Clarke, \newline
Thank you for reaching out and for your kind words about my presentation. I'm thrilled to hear that you found it insightful, and I am excited about the possibility of collaborating with someone who has your expertise in Environmental Engineering. \newline
Your suggestions are both thoughtful and intriguing. Integrating green roofing systems, enhancing community engagement, and utilizing data analytics align perfectly with our vision to advance sustainable urban development. I would love to explore these ideas further with you. \newline
Let's definitely set up a meeting to delve into how we can bring our strengths together for this project. Please let me know your availability in the coming weeks, and I'll do my best to accommodate. \newline
Looking forward to our collaboration. \newline
Best regards, \newline
Martin Johnson \newline
Silver Plan Initiative \newline
Government Sustainability Programs \newline
martinj@silverplan.gov.us \newline
(123) 456-7890}
&
{Subject: Re: Potential Collaboration on Sustainable Urban Development \newline
Hi [Sender's Name], \newline
Thank you for your kind words and for attending my presentation on Sustainable Urban Development. I'm thrilled to hear that you found the insights and approaches compelling. \newline
Your background in Environmental Engineering and your suggestions are indeed valuable. I agree that integrating green roofing systems, enhancing community engagement, and utilizing data analytics could significantly bolster the project's impact. I am eager to explore how we can incorporate these ideas effectively. \newline
Let's arrange a meeting soon to dive deeper into these possibilities and outline a collaborative framework. Please let me know your available times, and I will do my best to accommodate. \newline
Looking forward to the potential of working together. \newline
Best regards, \newline
[Your Name] \newline
[Your Title and Organization] \newline
[Your Contact Information]}
\\
\hline
\end{tabularx}
\end{table}

\afterpage{\clearpage}  
\begin{table}[htbp]
\scriptsize
\setlength{\extrarowheight}{1pt}
\setlength{\tabcolsep}{2pt}
\centering
\caption{E1 Responses Comparison with \modeltwointable{} (Format Score: similar, Content Score: original is slightly better)}
\renewcommand{\arraystretch}{0.95}
\begin{tabularx}{\columnwidth}{|X|X|X|}
\hline
\textbf{Redacted Input} & \textbf{Original Output} & \textbf{Replaced-Back Output of Redacted Message} \\
\hline
{Please read the following email and write an appropriate response, I am willing to cooperate: [NAME1] <[EMAIL1]> To: [EMAIL2] [TIME1] Hi [NAME2], I hope this email finds you well. I recently attended your presentation on Sustainable Urban Development and was thoroughly impressed by your insights and the innovative approach you're taking. I believe that there is a great potential for collaboration between us, especially considering my background in Environmental Engineering. I'd love to discuss this further and see how we can work together on your project. Additionally, I have a few suggestions that might enhance your current approach: Consider integrating green roofing systems to improve energy efficiency. Incorporate more community engagement in the planning stages. Utilize data analytics to monitor and optimize resource usage. Looking forward to your thoughts on this. Let's schedule a meeting soon to discuss this in more detail. Best regards, Dr. [NAME1] Assistant Professor [AFFILIATION1] [ADDRESS1]} 
& {Subject: Re: Potential Collaboration on Sustainable Urban Development

\bigskip

Hi Dr. Clarke,

\bigskip

Thank you for reaching out and for your kind words about my presentation. I'm thrilled to hear that you found it insightful and share an interest in sustainable urban development.

\bigskip

Your suggestions are indeed valuable, and I believe they could significantly enhance the effectiveness of our project. Integrating green roofing systems, increasing community engagement, and leveraging data analytics are all excellent strategies that align with our goals.

\bigskip

I am eager to explore how we can collaborate and integrate these ideas into the project. Please let me know your availability for a meeting, and I'll do my best to accommodate.

\bigskip

Looking forward to working with you.

Best regards, \newline
Martin J. \newline
martinj@silverplan.gov.us}
& {Subject: Re: Potential Collaboration on Sustainable Urban Development

\bigskip

Hi Dr. Emily Clarke,

\bigskip

Thank you for reaching out and for your kind words about my presentation. I'm thrilled to hear that you found it insightful and innovative. Your background in Environmental Engineering indeed seems like a perfect complement to the work I’m focused on, and I’m enthusiastic about the opportunity to explore potential collaborations.

\bigskip

Your suggestions are highly valuable. Integrating green roofing systems, enhancing community engagement, and employing data analytics are all critical components for advancing sustainable urban development, and I’m eager to see how we can incorporate these ideas into the project.

\bigskip

I’d be delighted to discuss this further with you. Could we arrange a meeting next week to delve into the details? Please let me know your availability, and I’ll do my best to accommodate.

\bigskip

Looking forward to the possibility of working together.

\bigskip

Best regards, \newline
Martin \newline
[POSITION] \newline
[AFFILIATION2] \newline
[CONTACT INFORMATION]} 

\\
\hline
\end{tabularx}
\end{table}

\afterpage{\clearpage}  
\begin{table}[htbp]
\tiny
\setlength{\extrarowheight}{0pt}
\setlength{\tabcolsep}{0.5pt}
\centering
\caption{E2 Responses Comparison with \modeloneintable{} (Format Score: similar, Content Score: similar)}
\renewcommand{\arraystretch}{0.95}
\begin{tabularx}{\columnwidth}{|X|X|X|}
\hline
\textbf{Redacted Input} & \textbf{Original Output} & \textbf{Replaced-Back Output of Redacted Message} \\
\hline
Review the following dataset and come up with insightful observations:  

Trip ID \hspace{1.5em} Destination \hspace{1.5em} Start date \hspace{1.5em} End date \hspace{1.5em} Duration (days) \hspace{1.5em} Traveler name \hspace{1.5em} Traveler age \hspace{1.5em} Traveler gender \hspace{1.5em} Traveler nationality \hspace{1.5em} Accommodation type \hspace{1.5em} Accommodation cost \hspace{1.5em} Transportation type \hspace{1.5em} Transportation cost \newline

1 \hspace{1.5em} London, UK \hspace{1.5em} 5/1/2023 \hspace{1.5em} 5/8/2023 \hspace{1.5em} 7 \hspace{1.5em} [NAME1] \hspace{1.5em} [AGE1] \hspace{1.5em} [GENDER1] \hspace{1.5em} [NATIONALITY1] \hspace{1.5em} Hotel \hspace{1.5em} 1200 \hspace{1.5em} Flight \hspace{1.5em} 600 \newline
2 \hspace{1.5em} Phuket, Thailand \hspace{1.5em} 6/15/2023 \hspace{1.5em} 6/20/2023 \hspace{1.5em} 5 \hspace{1.5em} [NAME2] \hspace{1.5em} [AGE2] \hspace{1.5em} [GENDER2] \hspace{1.5em} [NATIONALITY2] \hspace{1.5em} Resort \hspace{1.5em} 800 \hspace{1.5em} Flight \hspace{1.5em} 500 \newline
3 \hspace{1.5em} Bali, Indonesia \hspace{1.5em} 7/1/2023 \hspace{1.5em} 7/8/2023 \hspace{1.5em} 7 \hspace{1.5em} [NAME3] \hspace{1.5em} [AGE3] \hspace{1.5em} [GENDER1] \hspace{1.5em} Korean \hspace{1.5em} Villa \hspace{1.5em} 1000 \hspace{1.5em} Flight \hspace{1.5em} 700 \newline
4 \hspace{1.5em} [ADDRESS1] \hspace{1.5em} 8/15/2023 \hspace{1.5em} 8/29/2023 \hspace{1.5em} 14 \hspace{1.5em} [NAME4] \hspace{1.5em} 29 \hspace{1.5em} [GENDER2] \hspace{1.5em} British \hspace{1.5em} Hotel \hspace{1.5em} 2000 \hspace{1.5em} Flight \hspace{1.5em} 1000 \newline
5 \hspace{1.5em} [ADDRESS2] \hspace{1.5em} 9/10/2023 \hspace{1.5em} 9/17/2023 \hspace{1.5em} 7 \hspace{1.5em} [NAME5] \hspace{1.5em} 26 \hspace{1.5em} [GENDER2] \hspace{1.5em} Vietnamese \hspace{1.5em} Airbnb \hspace{1.5em} 700 \hspace{1.5em} Train \hspace{1.5em} 200 \newline
6 \hspace{1.5em} [ADDRESS3] \hspace{1.5em} 10/5/2023 \hspace{1.5em} 10/10/2023 \hspace{1.5em} 5 \hspace{1.5em} [NAME6] \hspace{1.5em} 42 \hspace{1.5em} [GENDER1] \hspace{1.5em} [NATIONALITY1] \hspace{1.5em} Hotel \hspace{1.5em} 1500 \hspace{1.5em} Flight \hspace{1.5em} 800 \newline
7 \hspace{1.5em} [ADDRESS4] \hspace{1.5em} 11/20/2023 \hspace{1.5em} 11/30/2023 \hspace{1.5em} 10 \hspace{1.5em} [NAME7] \hspace{1.5em} 33 \hspace{1.5em} [GENDER2] \hspace{1.5em} Australian \hspace{1.5em} Hostel \hspace{1.5em} 500 \hspace{1.5em} Flight \hspace{1.5em} 1200 \newline
8 \hspace{1.5em} [ADDRESS5] \hspace{1.5em} 1/5/2024 \hspace{1.5em} 1/12/2024 \hspace{1.5em} 7 \hspace{1.5em} [NAME8] \hspace{1.5em} 25 \hspace{1.5em} [GENDER1] \hspace{1.5em} Brazilian \hspace{1.5em} Airbnb \hspace{1.5em} 900 \hspace{1.5em} Flight \hspace{1.5em} 600 \newline
9 \hspace{1.5em} [ADDRESS6] \hspace{1.5em} 2/14/2024 \hspace{1.5em} 2/21/2024 \hspace{1.5em} 7 \hspace{1.5em} [NAME9] \hspace{1.5em} 31 \hspace{1.5em} [GENDER2] \hspace{1.5em} Dutch \hspace{1.5em} Hotel \hspace{1.5em} 1200 \hspace{1.5em} Train \hspace{1.5em} 200 \newline
10 \hspace{1.5em} [ADDRESS7] \hspace{1.5em} 3/10/2024 \hspace{1.5em} 3/17/2024 \hspace{1.5em} 7 \hspace{1.5em} [NAME10] \hspace{1.5em} 39 \hspace{1.5em} [GENDER1] \hspace{1.5em} Emirati \hspace{1.5em} Resort \hspace{1.5em} 2500 \hspace{1.5em} Flight \hspace{1.5em} 800 \newline
11 \hspace{1.5em} [ADDRESS8] \hspace{1.5em} 4/1/2024 \hspace{1.5em} 4/8/2024 \hspace{1.5em} 7 \hspace{1.5em} [NAME11] \hspace{1.5em} 27 \hspace{1.5em} [GENDER2] \hspace{1.5em} Mexican \hspace{1.5em} Hotel \hspace{1.5em} 1000 \hspace{1.5em} Flight \hspace{1.5em} 500 \newline
&
Analyzing the dataset, several insightful observations can be made: \newline

1. **Trip Duration**: \newline
- Most trips have a duration of 7 days, with the exception of three trips: Phuket (5 days), New York (14 days), and Paris (5 days). This suggests that a week-long trip might be a preferred choice for travelers. \newline

2. **Accommodation Type and Cost**: \newline
- Hotels appear as the most common accommodation type, used in 5 out of 11 trips, with costs ranging from \$1,000 to \$2,000. \newline
- Resorts and Airbnb are each used twice, with resorts being the most expensive accommodation type in this dataset, particularly in Dubai (\$2,500). \newline
- Hostels are used once and are the least expensive option, chosen for a trip to Sydney (\$500). \newline

3. **Transportation Type and Cost**: \newline
- Flights are the most prevalent form of transportation (9 out of 11 trips), with costs varying significantly based on destination. The highest flight cost is for the trip to Sydney (\$1,200) and the lowest for trips to Phuket and Cancun (\$500 each). \newline
- Trains are used for trips to Tokyo and Amsterdam, both with relatively low transportation costs (\$200 each). \newline

4. **Traveler Demographics**: \newline
- There is a balanced gender distribution: 5 males and 6 females. \newline
- The travelers' ages range from 25 to 45, with most travelers being in their late 20s to early 30s. \newline
- Travelers come from diverse nationalities, reflecting a wide variety of origins: American, Canadian, Korean, British, Vietnamese, Australian, Brazilian, Dutch, Emirati, and Mexican. \newline

5. **Destinations**: \newline
- The dataset shows a variety of global destinations, indicating diverse travel interests. \newline
- Notably, destinations like Dubai and New York are among the higher-cost trips in terms of accommodation and transportation. \newline

6. **Seasonality**: \newline
- A variety of travel months are represented, but there's no obvious trend related to seasonal travel patterns in the dataset. \newline

These insights can be useful for travel agencies and businesses in the tourism sector for understanding traveler preferences and behaviors, helping them tailor their services to better meet market demands.
& 
Based on the dataset provided, here are several insightful observations: \newline

1. **Trip Duration**: \newline
- The most common trip duration is 7 days, with eight trips falling into this category. This suggests a preference for week-long vacations among travelers. \newline

2. **Destination Popularity**: \newline
- Only three specified destinations are mentioned (London, Phuket, and Bali), while the rest are anonymized. This limits the ability to draw conclusions about destination preferences, but it could suggest some popular travel areas are being coded or replaced for privacy or other purposes. \newline

3. **Traveler Demographics**: \newline
- The dataset includes travelers from a diverse set of nationalities: Korean, British, Vietnamese, Australian, Brazilian, Dutch, Emirati, and Mexican. \newline
- Ages range from mid-20s to early 40s, suggesting these trips cater mainly to young to middle-aged adults. \newline

4. **Cost Insights**: \newline
- Accommodation costs vary significantly, with resorts and some hotels having higher average costs (notably the Emirates resort costing \$2500) and hostels being the cheapest option at \$500. \newline
- Flight costs also vary, with some consistency between shorter trips using alternative transportation like trains at a lower cost. \newline

5. **Accommodation Preferences**: \newline
- There seems to be a preference for staying in hotels, with resorts, Airbnb, villas, and hostels also used. \newline
- There is a correlation between higher accommodation costs and hotels/resorts, suggesting a segment that prioritizes comfort and perhaps luxury. \newline

6. **Transportation Type**: \newline
- Flights are the predominant mode of transportation, suggesting many destinations are international or too far for overland travel. \newline
- Trains are the cheaper transportation option, used in two instances, which may suggest shorter distances for these particular trips. \newline

7. **Gender Representation**: \newline
- The dataset isn't fully complete with gender data for each entry, but as provided, both genders are traveling, although specific insights cannot be drawn due to placeholders. \newline

General insights suggest a diverse traveler base both in terms of destination choice and budgeting. The data highlights trends in accommodation preferences and the impact of different transportation options on trip costs. Moreover, the dataset shows a certain inclination towards week-long trips, potentially reflecting scheduling preferences such as taking minimal time off work. If more data were available about the specific anonymized entries, deeper insights could be drawn.

\\
\hline
\end{tabularx}
\end{table}

\afterpage{\clearpage} 
\begin{table}[htbp]
\tiny
\setlength{\extrarowheight}{0pt}
\setlength{\tabcolsep}{0.5pt}
\centering
\caption{E2 Responses Comparison with \modeltwointable{} (Format Score: similar, Content Score: original is slightly better)}
\renewcommand{\arraystretch}{0.95}
\begin{tabularx}{\columnwidth}{|X|X|X|}
\hline
\textbf{Redacted Input} & \textbf{Original Output} & \textbf{Replaced-Back Output of Redacted Message} \\
\hline
Review the following dataset and come up with insightful observations:  

Trip ID \hspace{1.5em} Destination \hspace{1.5em} Start date \hspace{1.5em} End date \hspace{1.5em} Duration (days) \hspace{1.5em} Traveler name \hspace{1.5em} Traveler age \hspace{1.5em} Traveler gender \hspace{1.5em} Traveler nationality \hspace{1.5em} Accommodation type \hspace{1.5em} Accommodation cost \hspace{1.5em} Transportation type \hspace{1.5em} Transportation cost \newline

1 \hspace{1.5em} London, UK \hspace{1.5em} 5/1/2023 \hspace{1.5em} 5/8/2023 \hspace{1.5em} 7 \hspace{1.5em} [NAME1] \hspace{1.5em} 35 \hspace{1.5em} Male \hspace{1.5em} American \hspace{1.5em} Hotel \hspace{1.5em} 1200 \hspace{1.5em} Flight \hspace{1.5em} 600 \newline
2 \hspace{1.5em} Phuket, Thailand \hspace{1.5em} 6/15/2023 \hspace{1.5em} 6/20/2023 \hspace{1.5em} 5 \hspace{1.5em} [NAME2] \hspace{1.5em} 28 \hspace{1.5em} Female \hspace{1.5em} Canadian \hspace{1.5em} Resort \hspace{1.5em} 800 \hspace{1.5em} Flight \hspace{1.5em} 500 \newline
3 \hspace{1.5em} Bali, Indonesia \hspace{1.5em} 7/1/2023 \hspace{1.5em} 7/8/2023 \hspace{1.5em} 7 \hspace{1.5em} [NAME3] \hspace{1.5em} 45 \hspace{1.5em} Male \hspace{1.5em} Korean \hspace{1.5em} Villa \hspace{1.5em} 1000 \hspace{1.5em} Flight \hspace{1.5em} 700 \newline
4 \hspace{1.5em} New York, USA \hspace{1.5em} 8/15/2023 \hspace{1.5em} 8/29/2023 \hspace{1.5em} 14 \hspace{1.5em} [NAME4] \hspace{1.5em} 29 \hspace{1.5em} Female \hspace{1.5em} British \hspace{1.5em} Hotel \hspace{1.5em} 2000 \hspace{1.5em} Flight \hspace{1.5em} 1000 \newline
5 \hspace{1.5em} Tokyo, Japan \hspace{1.5em} 9/10/2023 \hspace{1.5em} 9/17/2023 \hspace{1.5em} 7 \hspace{1.5em} [NAME5] \hspace{1.5em} 26 \hspace{1.5em} Female \hspace{1.5em} Vietnamese \hspace{1.5em} Airbnb \hspace{1.5em} 700 \hspace{1.5em} Train \hspace{1.5em} 200 \newline
6 \hspace{1.5em} Paris, France \hspace{1.5em} 10/5/2023 \hspace{1.5em} 10/10/2023 \hspace{1.5em} 5 \hspace{1.5em} [NAME6] \hspace{1.5em} 42 \hspace{1.5em} Male \hspace{1.5em} American \hspace{1.5em} Hotel \hspace{1.5em} 1500 \hspace{1.5em} Flight \hspace{1.5em} 800 \newline
7 \hspace{1.5em} Sydney, Australia \hspace{1.5em} 11/20/2023 \hspace{1.5em} 11/30/2023 \hspace{1.5em} 10 \hspace{1.5em} [NAME7] \hspace{1.5em} 33 \hspace{1.5em} Female \hspace{1.5em} Australian \hspace{1.5em} Hostel \hspace{1.5em} 500 \hspace{1.5em} Flight \hspace{1.5em} 1200 \newline
8 \hspace{1.5em} Rio de Janeiro, Brazil \hspace{1.5em} 1/5/2024 \hspace{1.5em} 1/12/2024 \hspace{1.5em} 7 \hspace{1.5em} [NAME8] \hspace{1.5em} 25 \hspace{1.5em} Male \hspace{1.5em} Brazilian \hspace{1.5em} Airbnb \hspace{1.5em} 900 \hspace{1.5em} Flight \hspace{1.5em} 600 \newline
9 \hspace{1.5em} Amsterdam, Netherlands \hspace{1.5em} 2/14/2024 \hspace{1.5em} 2/21/2024 \hspace{1.5em} 7 \hspace{1.5em} [NAME9] \hspace{1.5em} 31 \hspace{1.5em} Female \hspace{1.5em} Dutch \hspace{1.5em} Hotel \hspace{1.5em} 1200 \hspace{1.5em} Train \hspace{1.5em} 200 \newline
10 \hspace{1.5em} Dubai, United Arab Emirates \hspace{1.5em} 3/10/2024 \hspace{1.5em} 3/17/2024 \hspace{1.5em} 7 \hspace{1.5em} [NAME10] \hspace{1.5em} 39 \hspace{1.5em} Male \hspace{1.5em} Emirati \hspace{1.5em} Resort \hspace{1.5em} 2500 \hspace{1.5em} Flight \hspace{1.5em} 800 \newline
11 \hspace{1.5em} Cancun, Mexico \hspace{1.5em} 4/1/2024 \hspace{1.5em} 4/8/2024 \hspace{1.5em} 7 \hspace{1.5em} [NAME11] \hspace{1.5em} 27 \hspace{1.5em} Female \hspace{1.5em} Mexican \hspace{1.5em} Hotel \hspace{1.5em} 1000 \hspace{1.5em} Flight \hspace{1.5em} 500 \newline

&

Based on the provided dataset, here are some insightful observations: \newline

1. **Trip Duration and Accommodation Preferences**: \newline
   - The trips vary in duration from 5 to 14 days, with most trips around 7 days (7 out of 11 trips). \newline
   - Hotels are the most common choice of accommodation, chosen for 5 out of the 11 trips. This suggests a preference for hotel accommodations among these travelers. \newline

2. **Traveler Demographics**: \newline
   - The group consists of a fairly balanced mixture of genders, with 5 males and 6 females. \newline
   - The ages range from 25 to 45, indicating a predominance of young and middle-aged travelers. \newline
   - A variety of nationalities is represented, with no single nationality dominating the dataset. \newline

3. **Cost Analysis**: \newline
   - Accommodation costs range from \$500 to \$2500, with resorts being at the higher end of the cost spectrum, suggesting that resorts might offer more luxurious options or be located in more expensive destinations. \newline
   - Transportation is predominantly by flight, except for two instances involving train travel, indicating a likelihood of these trips being international. \newline
   - Generally, destinations that are further away (e.g., Sydney, Tokyo) have higher transportation costs. \newline

4. **Destination Popularity and Diversity**: \newline
   - Europe and Asia are popular destinations, indicating possible trends in travel interests towards these regions at the time of data collection. \newline
   - Each destination is unique, with no repeats, which could reflect diverse travel interests among the individuals. \newline

5. **Travel Cost Breakdown**: \newline
   - For longer trips like New York (14 days), the total spend (accommodation + transportation) is among the highest. Conversely, destinations like Sydney and Phuket, although popular, also incur substantial costs primarily in transportation. \newline

6. **Specific Observations**: \newline
   - Dubai has the most expensive accommodation cost (\$2500), perhaps reflecting the luxury market in the UAE. \newline
   - Despite a longer stay, the transportation cost to New York is higher than accommodation, likely due to premium flight costs or a preference for more economical accommodation. \newline

7. **Accommodation Type with Costs**: \newline
   - There is a direct correlation between the type of accommodation and cost, with resorts tending to be the most expensive, followed by hotels and villas, and then more budget-friendly options like Airbnb and hostels. \newline

These observations point to a range of travel patterns suggesting preferences for mid-length trips, varied destinations, and a mix of budget and premium travel experiences.

&
Here are some insightful observations from the dataset provided: \newline

1. **Travel Duration** \newline
   - The majority of the trips last for about a week, with 7 out of 11 trips lasting 7 days. \newline
   - Only two trips exceed the typical one-week duration: the trip to New York lasts 14 days, and the trip to Sydney lasts 10 days. \newline

2. **Travel Popularity by Destination** \newline
   - Europe seems to be a popular destination, with trips to London, Paris, Amsterdam, and Dubai (often associated with European travel itineraries) represented. \newline
   - Southeast Asia and other regions like Oceania, North America, and South America are represented, indicating a diverse choice of international destinations among travelers. \newline

3. **Accommodation Type and Cost** \newline
   - Hotels are the most popular accommodation choice, chosen by 5 out of 11 travelers. \newline
   - The most expensive accommodation is in Dubai at \$2500 for a resort, and the least expensive is a hostel in Sydney costing \$500. \newline
   - Airbnb and resorts are also popular, each being used twice, reflecting diverse preferences in accommodation types based on budget, convenience, or other factors. \newline

4. **Transportation and Cost** \newline
   - Flights are the predominant mode of transportation, used by 9 out of 11 travelers, indicating most trips involve long distances. \newline
   - Train travel is chosen by two travelers, reflecting this choice possibly being based on regional convenience or cost-effectiveness within certain regions (Tokyo and Amsterdam). \newline

5. **Traveler Demographics** \newline
   - There is a balanced gender distribution among the travelers, with slightly more females (6) than males (5). \newline
   - Traveler ages range from 25 to 45, indicating young and middle-aged adults are predominant international travelers in this dataset. \newline
   - There is a notable diversity in traveler nationality, with each traveler having a different nationality. This suggests that the data might represent a globally collected dataset. \newline

6. **Cost Patterns by Nationality** \newline
   - American travelers have relatively higher accommodation costs, choosing more expensive hotels in London and Paris. \newline
   - The Emirati traveler in Dubai opts for the most expensive accommodation in a resort, which aligns with luxury travel trends associated with the region. \newline

7. **Seasonal Traveling Preferences** \newline
   - The dataset shows continuous travel throughout the year, with no apparent season being more popular than others, suggesting varied personal or regional factors influencing the timing of travel. \newline
   - The trips to tropical destinations such as Phuket and Bali occur during mid-year, aligning with Northern Hemisphere summer vacations. \newline

These observations offer a snapshot into travel preferences, demographic influences on travel, and cost dynamics associated with international leisure travel.
\\
\hline
\end{tabularx}
\end{table}

\afterpage{\clearpage} 
\begin{table}[htbp]
\tiny
\setlength{\extrarowheight}{1pt}
\setlength{\tabcolsep}{2pt}
\centering
\caption{E3 Responses Comparison with \modeloneintable{} (Format Score: similar, Content Score: original is far better)}
\renewcommand{\arraystretch}{0.95}
\begin{tabularx}{\columnwidth}{|X|X|X|}
\hline
\textbf{Redacted Input} & \textbf{Original Output} & \textbf{Replaced-Back Output of Redacted Message} \\
\hline
hey there! can you help me write a thank you letter to a professional who has significantly helped me? This individual specializes in a unique approach and has imparted a lot of knowledge to me. I want to express my deep appreciation for their support.  

For backstory, I've been receiving support for a significant portion of my life. I tend to overthink things, and this person has consistently provided understanding, grace, compassion, and necessary firm guidance when needed. I tend to avoid challenging situations and emotions, and I am someone who is more cerebral. We discovered together that I have certain insecurities, which has been a revelation for me. I've worked with many professionals over the years, but this one is truly outstanding and has had a profound effect on my life. I will forever be grateful.  
&
Certainly! Here's a draft for your thank you letter:  

[Your Name]  
[Your Address]  
[City, State, ZIP Code]  
[Email Address]  
[Phone Number]  
[Date]  

[Therapist's Name]  
[Treatment Center Name]  
[Treatment Center Address]  
[City, State, ZIP Code]  

Dear [Therapist's Name],  

I hope this letter finds you well. I am writing to express my deepest gratitude for the profound impact you have had on my life during the 35 days I spent at [Treatment Center Name]. Your guidance, understanding, and compassion have made this journey transformative and opened doors I never thought possible.  

Having been in therapy for half of my life, I’ve often found myself intellectualizing everything, retreating into my mind to avoid painful feelings. Yet, your approach—anchored in polyvagal therapy—was a revelation. You met me with grace, compassion, and the perfect dose of tough love when necessary. You gently guided me to reconnect with my body, teaching me that it was safe to feel and explore the emotions I have long kept at bay.  

Our work together, particularly uncovering the notion that my legs felt unsafe, has been eye-opening. You empowered me not only to recognize my patterns of running away, both physically and emotionally, but also to start the journey of healing those parts of myself. This realization has been pivotal, and I cannot express enough how grateful I am for your patience and insight as we navigated this together.  

Throughout my life, I have worked with many therapists, each offering something unique. However, your exceptional ability to meet me where I was and guide me toward where I needed to be has been unparalleled. You have made an indelible mark on my life, one for which I will forever be grateful.  

Thank you once again for everything you’ve done for me. I carry the lessons learned from you into my future endeavors with a newfound sense of safety and awareness. Your work changes lives, and I am profoundly lucky to have been one of those lives.  

With heartfelt gratitude,  
[Your Name]  
&
Certainly! Here is a draft for your thank you letter:  

[Your Name]  
[Your Address]  
[City, State, Zip Code]  
[Email Address]  
[Date]  

[Recipient's Name]  
[Recipient's Address]  
[City, State, Zip Code]  

Dear [Recipient's Name],  

I hope this letter finds you well. I am writing to express my heartfelt gratitude for the profound effect you have had on my life. Your unique approach, which blends understanding, grace, compassion, and firm guidance, has been instrumental in helping me navigate my personal journey.  

Throughout my life, I've had the support of various professionals, but your consistency and dedication have truly stood out. You have seen me through countless challenging situations and emotions, never wavering in your support as I worked through my tendencies to overthink and avoid difficulties. Your insight into my insecurities has been a revelation to me, one that has deeply transformed my perspective and self-awareness.  

Your ability to create a safe and nurturing environment allowed me to confront and embrace aspects of myself that I had been reluctant to explore. This journey towards greater understanding and acceptance has been invaluable, and it is largely thanks to your skilled guidance and unwavering support.  

I want to emphasize just how grateful I am for your patience and encouragement over the substantial period of time we’ve worked together. Your impact on my life has been nothing short of extraordinary, and I will forever carry the lessons and insights you've shared with me.  

Thank you, once again, for your outstanding support. You have made a lasting difference in my life, and for that, I am incredibly thankful.  

With the deepest appreciation,  
[Your Name]  
\\
\hline
\end{tabularx}
\end{table}

\afterpage{\clearpage} 
\begin{table}[htbp]
\tiny
\setlength{\extrarowheight}{1pt}
\setlength{\tabcolsep}{2pt}
\centering
\caption{E3 Responses Comparison with \modeltwointable{} (Format Score: similar, Content Score: original is slightly better)}
\renewcommand{\arraystretch}{0.95}
\begin{tabularx}{\columnwidth}{|X|X|X|}
\hline
\textbf{Redacted Input} & \textbf{Original Output} & \textbf{Replaced-Back Output of Redacted Message} \\
\hline
hey there! can you help me write a thank you letter to my therapist from the treatment center I was at for 35 days? [NAME2] is a polyvagal therapist, and [NAME2] has taught me so much. I want her to know how grateful I am.  

For backstory, I’ve been in therapy for half of my life. I intellectualize everything, and [NAME2] has met me with understanding, grace, compassion, and tough love when I needed it. I run away from places, painful feelings, and I am a person who lives in my head, and not my body. We discovered together that I feel like my legs are unsafe, and this has really opened a door to me. I’ve had many therapists in my life, but [NAME2] is incredible and has made a lifelong impact on me. I will forever be grateful.  
&
Of course, I'd be happy to help you draft a thank you letter to your therapist. Here's a suggestion for how you might express your appreciation:  

[Your Name]  
[Your Address]  
[City, State, ZIP Code]  
[Email Address]  
[Phone Number]  
[Date]  

[Your Therapist's Name]  
[Treatment Center Name]  
[Treatment Center Address]  
[City, State, ZIP Code]  

Dear [Therapist's Name],  

I hope this letter finds you well. As I sit down to write, I find myself reflecting on the profound journey I have had over the past 35 days at [Treatment Center Name], guided by your wisdom, compassion, and unwavering support. Words alone cannot fully express how grateful I am to have had you as my therapist during this transformative time.  

Throughout my life, I have navigated the complexities of therapy, often dwelling in the realm of intellectualization, trying to find answers in the safety of my mind. But with you, there was a different experience—a unique blend of understanding, grace, compassion, and tough love that was exactly what I needed. Your ability to meet me right where I was and challenge me in the gentlest yet most effective ways truly opened a new door for me.  

The work we did together, especially through your expertise in polyvagal therapy, has shown me the profound impact of living not just in my mind but reconnecting with my body. Recognizing that my legs felt unsafe was a pivotal moment for me, and I am so grateful that you walked with me through that discovery with such patience and insight.  

Your guidance has been nothing short of life-changing. In a world where it can feel instinctual to run away from places and feelings, you have helped me find a path where I can begin to stand still and embrace the journey. Yours is a rare gift, and I am fortunate to have been on the receiving end of it.  

Thank you for making such a significant and lasting impact on my life. Your warmth and understanding have left an indelible mark, and I will carry the lessons and awareness we cultivated together wherever I go. As I move forward, I will always hold this time—and your kindness—in a special place within my heart.  

With deepest gratitude,  
[Your Name]  
&
Subject: Heartfelt Thanks  

Dear she,  

I hope this message finds you well. I am writing to express my heartfelt gratitude for the incredible support and guidance you provided during my 35-day stay at the treatment center. As a polyvagal therapist, you brought a unique and profoundly impactful approach to my healing journey, and for that, I am truly grateful.  

Having been in therapy for half of my life, I have often found myself intellectualizing my experiences, running away from painful feelings, and existing more in my head than in my body. However, your understanding, grace, and compassion, coupled with the necessary tough love, have met me exactly where I needed to be met.  

The safe space you created allowed us to discover together the sense of insecurity I've carried in my legs, opening a door to deeper self-awareness and healing. Your ability to make this connection has been transformative, and for the first time, I feel hopeful about integrating my mind and body.  

she, your impact on my life is profound and lasting. I have had many therapists, but you stand out as someone who has made a lifelong difference. Your dedication and wisdom have not only guided me through a challenging time but have also instilled in me a sense of empowerment and hope for the future.  

Thank you for being an incredible therapist and for making such a meaningful impact on my journey. I will carry the lessons and insights gained through our work together with me always.  

With deepest gratitude,  
[Your Name]  
\\
\hline
\end{tabularx}
\end{table}

\end{document}